\documentclass[noshowpacs,amsmath,twocolumn,superscriptaddress,10pt,aps,prb]{revtex4-1} 
\usepackage{setspace}
\usepackage{amsmath}
\usepackage{breqn}
\usepackage{graphicx}
\usepackage[nearskip,margin = 0pt,caption=false]{subfig}

\usepackage{verbatim}
\usepackage{amsfonts}
\usepackage{amssymb}
\usepackage{epstopdf}
\usepackage{lipsum}

\usepackage{siunitx}
\usepackage{xcolor}
\usepackage{array}
\usepackage[normalem]{ulem}
\usepackage[resetlabels,labeled]{multibib}
\DeclareGraphicsExtensions{.pdf,.eps,.png,.jpg,.mps}
\graphicspath{{SI_figures/}} 
\usepackage{etoolbox}
\usepackage{subfiles}
\usepackage{titletoc}
\contentsmargin{1.55em}
\dottedcontents{section}[3.8em]{}{2.3em}{1pc}
\dottedcontents{subsection}[6.1em]{}{3.2em}{1pc}
\DeclareMathOperator{\sech}{sech}
\newcommand{\dd}{\mathop{}\!\mathrm{d}}\usepackage{etoolbox}

\begin{document}
\title{Ultrabroadband Passive Laser Noise Suppression to Quantum Noise Limit \\ through on-chip Second Harmonic Generation}

\author{Geun Ho Ahn$^{1, 2, \dagger} $, 
    Ziyu Wang$^{1, \dagger}$, 
    Devin J. Dean$^{1}$, 
    Hubert S. Stokowski$^{1}$,
    Taewon Park$^{1, 2}$, 
    Martin M. Fejer$^{1}$, 
    Jonathan Simon$^{1, 3}$,  
    Amir H. Safavi-Naeini$^{1, \ast} $\\
\vspace{+0.05 in}
$^1$Department of Applied Physics and Ginzton laboratory, Stanford University, Stanford, California 94305, USA\\
$^2$Department of Electrical Engineering, Stanford University, Stanford, California 94305, USA\\
$^2$Department of Physics, Stanford University, Stanford, California 94305, USA\\
{\small $^{\ast}$ safavi@stanford.edu }}



\maketitle


\textbf{Laser intensity noise limits performance in quantum sensing, metrology, and computing. Existing stabilization methods face a trade-off between bandwidth and complexity: electronic feedback loops are speed-limited, while optical resonators are constrained by narrow linewidths and locking requirements. Here, we demonstrate an all-optical ``noise eater'' that passively suppresses intensity fluctuations from DC to $>$10 gigahertz. By leveraging high-efficiency second-harmonic generation in nanophotonic lithium niobate waveguides, we operate at a pump-depletion stationary point where input fluctuations are decoupled from the output to first order. This passive and nonresonant nanophotonic device suppresses relative intensity noise by 25 to 60 dB over the full measurement bandwidth and stabilizes a noisy fiber amplifier output to the shot-noise limit. Our results establish a scalable, wide-bandwidth paradigm for laser stabilization essential for high-throughput quantum technologies and deployable photonic sensing systems.}

Laser amplitude fluctuations are a pervasive limitation in precision sciences. Intensity noise reduces the signal-to-noise ratio in measurements, masking weak signals in optical sensing~\cite{casacio2021quantum_microscopy}, degrades atom-trapping lifetimes by parametric heating~\cite{atom_heating_noise, atom_heating_noise_NIST_2025}, and thereby limits performance in quantum computing~\cite{bluvstein2024logical}, metrology~\cite{LIGO_noise, LO_Noise_Requirement}, communications and optical-links~\cite{intensitynoise_optical_comm, RIN_in_optical_links}. In most precision applications, laser-intensity stabilization is employed to reduce intensity noise, thereby improving precision and performance. The key figures of merit for intensity-stabilization methods include the noise-suppression ratio, suppression bandwidth, insertion loss, and overall method complexity~\cite{hall_taubman_ye_book}.  

Active noise stabilizers based on electronic feedback loops driving electro-optic or acousto-optic modulators are widely used remedies, sensing instantaneous power and providing a correction to reduce excursions of the output power from the desired level~\cite{active_noise_eater_book, robertson1986intensity}. However, such active controls require high loop bandwidth, which poses challenges for broadband noise suppression~\cite{active_noise_eater_book}. Moreover, it is even more challenging to integrate the required discrete components of active stabilizers into a chip-scale platform.

Engineering internal dynamics of a laser via injection locking to narrow bandwidth feedback or filtering with a high finesse resonator has been shown to reduce phase noise and moderately reduce intensity noise~\cite{xiang2021high, nie2024turnkey}. However, the performance depends sensitively on the coupling conditions, back reflections, and resonator drift, and the high finesse of the resonator limits the achievable bandwidth~\cite{lihachev2022low}. Furthermore, the requirement to access and control the laser's internal dynamics limits this approach to a relatively narrow class of devices.

Optical nonlinearity offers a compelling route to intensity-noise reduction by exchanging noise between modes or dissipating it into an effective harmonic bath~\cite{siegman1962nonlinear, zia2025noise, SHG_Buffer_Solid_State_Laser_2015, inoue2002experimental}. While intracavity second harmonic generation (SHG) has demonstrated noise buffering~\cite{SHG_FH_Cavity_2024, SHG_1_5um_buffer}, reliance on high-finesse cavities fundamentally restricts the suppression bandwidth and necessitates complex locking. Traveling-wave interactions in single-pass devices can overcome these limitations, offering passive, broadband suppression~\cite{Porat_Freespace_Theory}. However, this approach requires exceptional nonlinear efficiency to operate at practical power levels. Integrated thin-film lithium niobate (TFLN) waveguides address this challenge, combining tight mode confinement with extended interaction lengths to enable highly efficient, scalable nonlinear devices~\cite{wang2018integrated, boes2023lithium, park2024single}.
\begin{figure*}[t!] 
	\centering
    \includegraphics[width=1\linewidth]{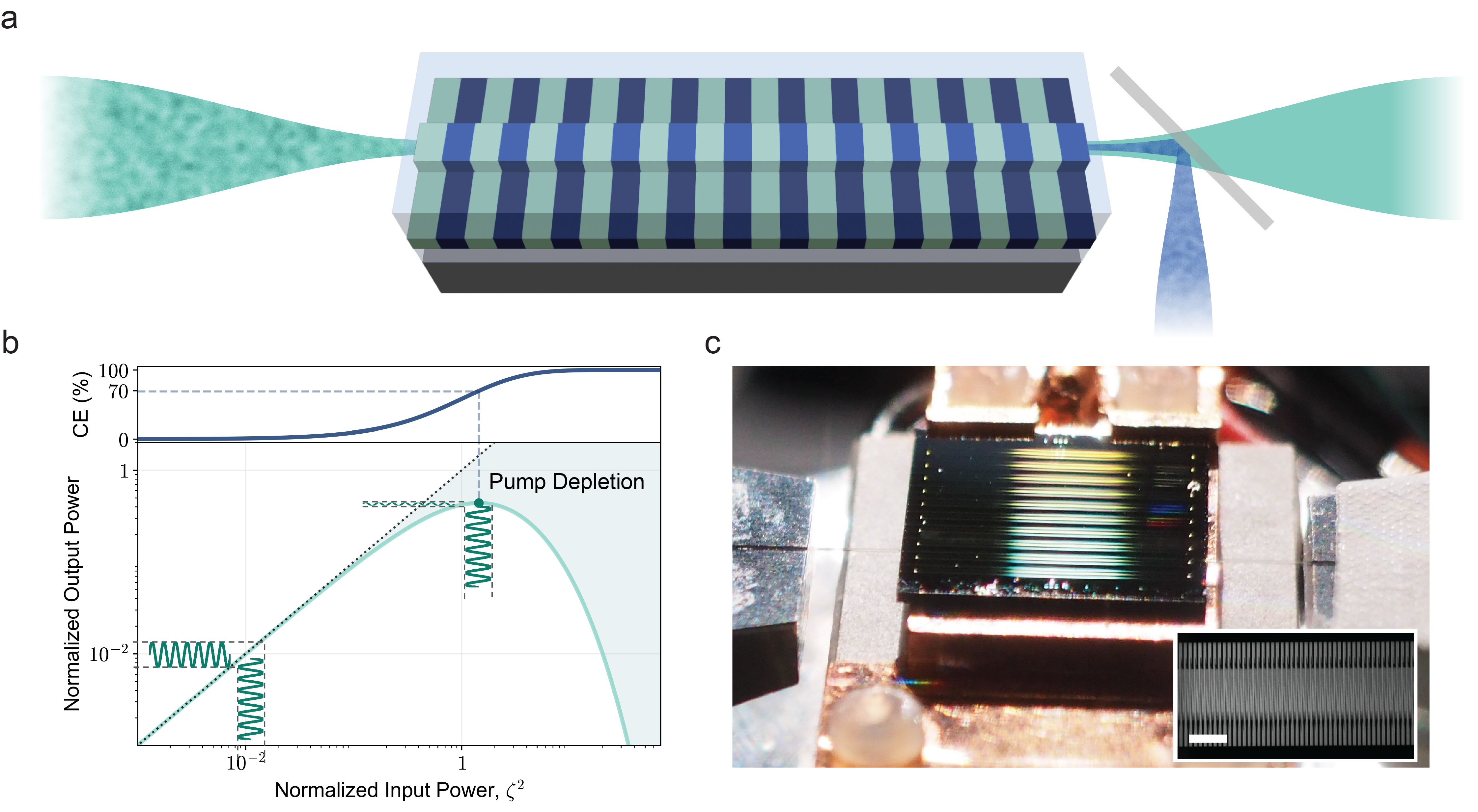}
    \caption{{\bf Theory of operation of PINE.} {\bf a}. Conceptual schematic showing the operation of the nonlinear optical noise eater, where a noisy input beam is coupled into PPTFLN and the noise on the output pump power is strongly suppressed at the critical point. The alternating colors of the PPTFLN conceptually represent inverted domains. {\bf b}. Simulated normalized first harmonic (FH) output power, which corresponds to output pump, and conversion efficiency (CE) into second harmonic (SH) as a function of normalized pump input power. At the CE $\approx 70\%$, a stationary point in normalized output power is observed. {\bf c}. An image of experimental setup with PINE chip, where input and output are coupled through lensed fibers. This $10$ mm $\times$ $10$ mm chip has 20 PINE devices, and each PINE section has a footprint around $0.05$ mm $\times$ $10$ mm. The inset image shows an SHG microscope image of periodically poled domains. The scale bar in the inset represents 30 $\mu m$. }
    \label{fig:Fig1}
\end{figure*}
Here, we demonstrate a passive and broadband noise-suppression method using a non-resonant SHG waveguide that is generalizable across wavelengths, power levels, and applications: At high conversion efficiency, with increasing input pump (FH) power, the depletion due to conversion to the SH increases, reaching a turning point where the output FH power begins to decrease~\cite{Porat_Freespace_Theory}. Operating at this stationary point where the input fluctuations are eliminated to first order, the photonic intensity noise eater (PINE) achieves $>$ 25 dB reduction in relative intensity noise (RIN), exhibits an ultra-wide suppression band ($>$ 10 GHz), and stabilizes the output of a noisy fiber amplifier to the standard quantum limit (SQL). To quantify the dependence on RF bandwidth, wavelength, and pump power, we develop new analytical and numerical sideband noise models for PINE that show excellent agreement with experiment. Compared with state-of-the-art active servo noise eaters, PINE eliminates feedback-loop bandwidth ceilings and actuator dynamic-range limits, remains compatible with a wide optical-power range and diverse wavelengths, and can be integrated upstream or downstream of other stabilization approaches. This stabilization technology, based on $\chi^{(2)}$, thus provides a new building block for intensity-stable laser systems suitable for atom- and ion- control, squeezed-light generation over audio-to-RF frequencies, and low-RIN sources for precision science and optical communications. 
\section{Theory of Operation}
\noindent
In guided wave systems, the SHG process is described by : 
    \begin{equation}\label{eq:1}
        \begin{aligned}
            \frac{\text{d} A}{\text{d} z}&=i\kappa A^*B e^{i\Delta k z}\\
            \frac{\text{d} B}{\text{d} z}&=i\kappa A^2 e^{-i\Delta k z}
        \end{aligned}
    \end{equation}
where $A$ and $B$ are the amplitudes of the pump and second harmonic fields normalized such that $|A|^2$, and $|B|^2$ are the power carried by the mode, $\kappa$ with units of $[\mathrm{W}^{-1/2}\mathrm{cm}^{-1}]$ is the second order nonlinear coefficient, and $\Delta k=k_B-2k_A-2\pi/\Lambda$ is the quasi-phase mismatch~\cite{jankowski2024ultrafast}, where $k_A$ and $k_B$ are the propagation constants of the modes, and $\Lambda$ is the period of the quasi-phase-matched (QPM) grating. For a phase matched interaction ($\Delta k = 0$), we analytically solve for $A$ and $B$ and find the corresponding relation between input and output powers:
\begin{equation}\label{eq:2}
    \begin{aligned}
        P_{p,\text{out}}&=|A(L)|^2=P_{p,\text{in}}\sech^2(\kappa \sqrt{P_{p,\text{in}}} L)\\
        P_{s,\text{out}}&=|B(L)|^2=P_{p,\text{in}}\tanh^2(\kappa \sqrt{P_{p,\text{in}}} L)\\
    \end{aligned}
\end{equation}
Using these solutions, we can relate the change in output pump power to a change in input pump power~\cite{Porat_Freespace_Theory}:
\begin{equation}\label{eq:3}
    \begin{aligned}
        \frac{\partial P_{p,\text{out}}}{\partial P_{p,\text{in}}}=\sech^2\left(\zeta \right)\left[1-\zeta \tanh{\left(\zeta \right)}\right]
    \end{aligned}
\end{equation}
\noindent where $\zeta = \kappa \sqrt{P_{p,\text{in}}} L$. Equation (\ref{eq:3}) shows that input power fluctuations are suppressed to first order when $\zeta\tanh{\zeta } = 1$ is satisfied\cite{Porat_Freespace_Theory}. This condition defines a critical pump power, $P_{\text{in, c}}$, at a critical value $\zeta_c \approx 1.2$. At this operating point, the system reaches a conversion efficiency (CE) of $\approx 70\%$, derived from the general depletion formula $P_{p,\text{out}}/P_{p,\text{in}} = 1- \text{CE} = \sech^2(\zeta)$. Consequently, the passive stabilization comes with a fixed intrinsic insertion loss of $-5.2$ dB, a universal value regardless of device length or efficiency.
\begin{figure}
\centering\includegraphics[width=\linewidth]{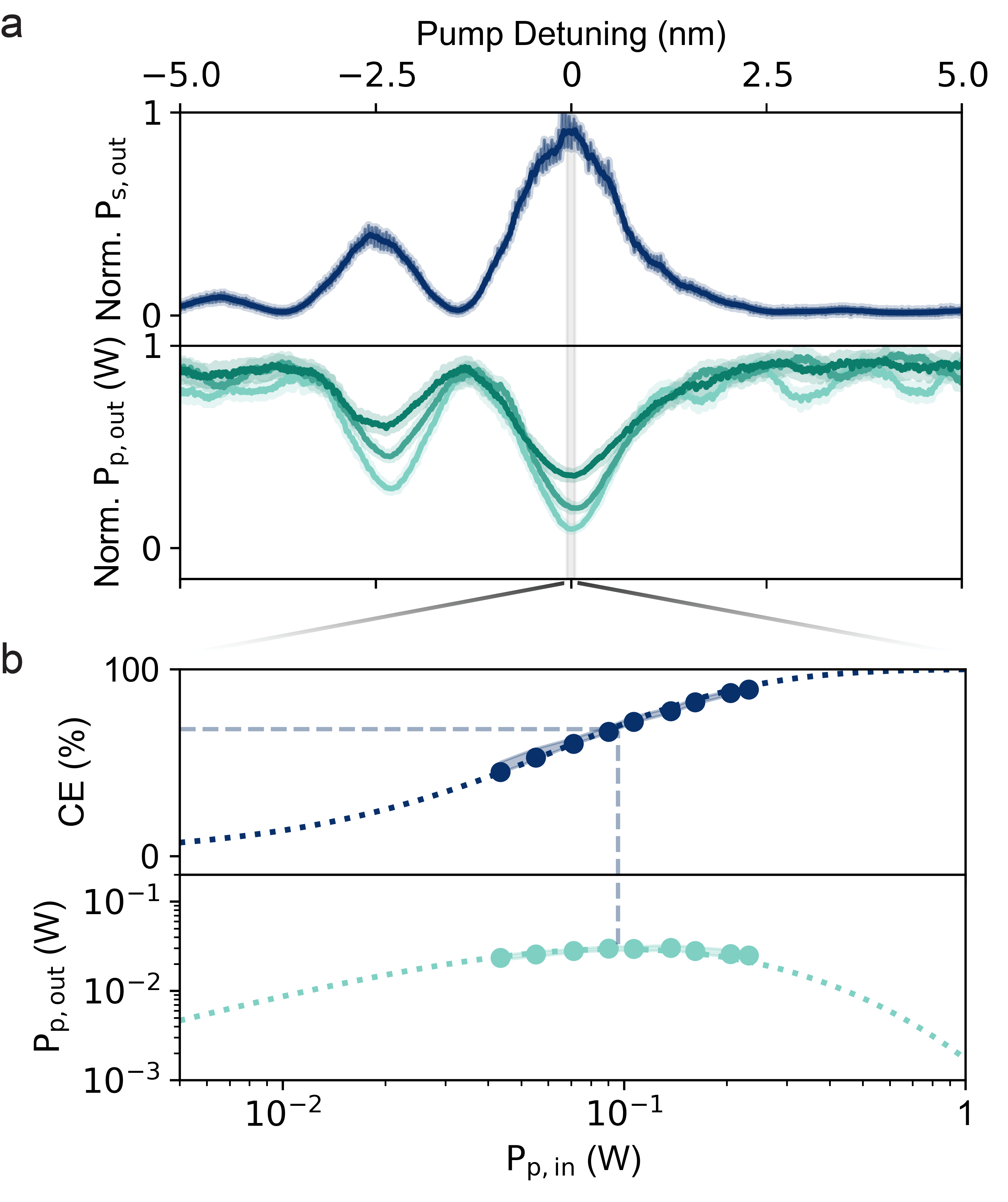}
\caption{{\bf Highly efficient integrated second harmonic generators.} {\bf a}. Normalized second harmonic generation intensity, Norm. $P_{s,\text{out}}$, (top panel) and normalized pump output, Norm. $P_{p,\text{out}}$, (bottom panel) as a function of pump wavelength detuning. The pump laser wavelength is swept around 1555.6 nm and the output pump (green) and second harmonic (blue) intensities measured, with a moving-average applied to the solid line, and the raw data plotted in fainter colors. For pump output scans, data from three different pump powers ($71.1$ mW, $137$ mW, $232$ mW) are plotted. At higher pump power, we observe lower normalized transmission at QPM wavelength, indicating greater pump depletion. The transfer function scan for SH was measured with $\approx 20$ mW of pump power on chip.  {\bf b}. Experimentally measured pump output power and CE into second harmonic as a function of input pump power at the QPM wavelength, indicated by the gray box in (a). The CE is estimated using pump depletion at QPM wavelength, following CE = $(P_{p, \text{in}} - P_{p,\text{out}})/P_{p, \text{in}} $. The dotted lines show analytical estimation of CE and pump depletion using the experimentally derived $\eta$ and $L$ = 10 mm} 
\label{fig:Fig2}
\end{figure}
SHG based on a guided mode enables high conversion efficiency and the pump depletion~\cite{parameswaran2002observation, chen2024adaptivepoling} required for  strong noise suppression at low optical pump powers. To demonstrate this nonlinear optical behavior, we explore the integrated $\chi^{(2)}$ waveguide setup shown in Fig.~\ref{fig:Fig1}a. A noisy input pump is coupled into a periodically poled integrated thin film lithium niobate (PPTFLN) waveguide. The waveguide generates a second-harmonic field which is filtered out.
Intuitively, as the input power increases along the pump-depletion curve in Fig. \ref{fig:Fig1}b, the output FH power first rises linearly and then flattens as the pump starts to be depleted. At the stationary point, $\zeta_c$, the slope ${\partial P_{p,\text{out}}}/{\partial P_{p,\text{in}}}=0$ so small input fluctuations are converted into changes in SH power rather than FH power, realizing a photonic integrated nonlinear noise eater.
An image of the fabricated PPTFLN waveguide chip in the measurement setup is shown in Fig.~\ref{fig:Fig1}c. The inset shows the periodically poled domains of the SHG waveguide imaged by second harmonic microscopy. To ensure high CE at low optical pump power, we use an adaptive poling technique for the PPTFLN waveguide~\cite{chen2024adaptivepoling}, with further fabrication details provided in the Methods and Fig. \ref{Sfig:Fab_Flow}. 

We characterize the SHG transfer function by scanning the pump wavelength near 1555.6 nm (Fig. \ref{fig:Fig2}a). The measured response exhibits a fidelity of $\approx 0.94$ relative to an ideal $\mathrm{sinc}^2$ profile, confirming the quality of the adaptive poling (see Methods and Fig. \ref{Sfig:CE_PD_Full}a~\cite{santandrea2019general_fidelity}). The SHG waveguide length is 10 mm, and by calibrating the input pump powers with the output pump powers and output SH powers, we estimate a peak efficiency $\eta =\kappa^2= 1500 \% \text{W}^{-1}\text{cm}^{-2}$. To study the input pump power dependence of pump depletion and CE, we repeat the transfer function measurements at different input pump powers. As input power increases, the normalized (fractional) pump output at the QPM wavelength decreases, indicating increased pump depletion (Fig.~\ref{fig:Fig2}a). A complete sweep of the input power dependence of pump depletion is shown in Fig. \ref{Sfig:CE_PD_Full} b. Fig.~\ref{fig:Fig2}b shows the experimentally measured dependence of CE and output pump power on input pump power. The dotted lines in each subplot correspond to an analytic estimate based on the Equation \ref{eq:2} with experimentally derived $\eta$ and the device length $L$. Scatter points represent the moving-average value of the transmission at QPM wavelength during the laser scan, to obviate the impact of a small standing wave from chip-edge reflections. The shaded area shows the minimum and maximum within the moving-average window. At the estimated critical power given the $\eta$ and $L$ ($P_{p, \text{in}}\approx 100 \text{ mW}$), Fig.~\ref{fig:Fig2}c shows the corresponding CE of $\approx 70 \%$, and the stationary point in the output power, indicating the zero noise transfer from the input to the output power.

\begin{figure}[h]
\centering\includegraphics[width=\linewidth]{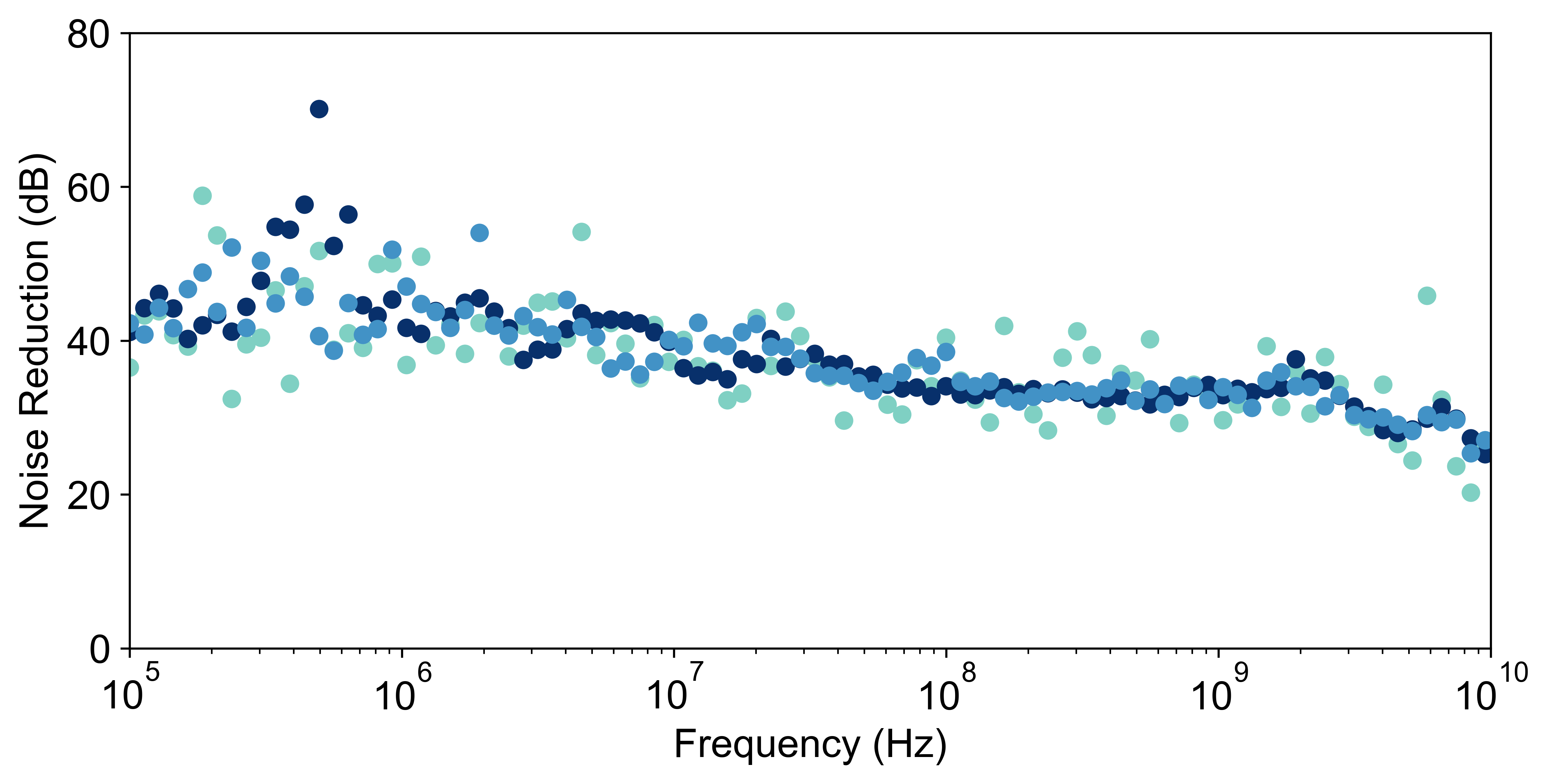}
\caption{{\bf Ultra-Broadband Intensity Noise Reduction.} {\bf a}. Experimental NRR over the broadband RF frequency of the VNA indicated by the scatter points. The measurement was performed with injected RIN peak values of -38.7 dBc/Hz (navey blue), -48.7 dBc/Hz (light blue), and -58.7 dBc/Hz (light green).}
\label{fig:Fig3}
\end{figure}
\section{Noise suppression bandwidth}
\begin{figure*}
\centering\includegraphics[width=1\linewidth]{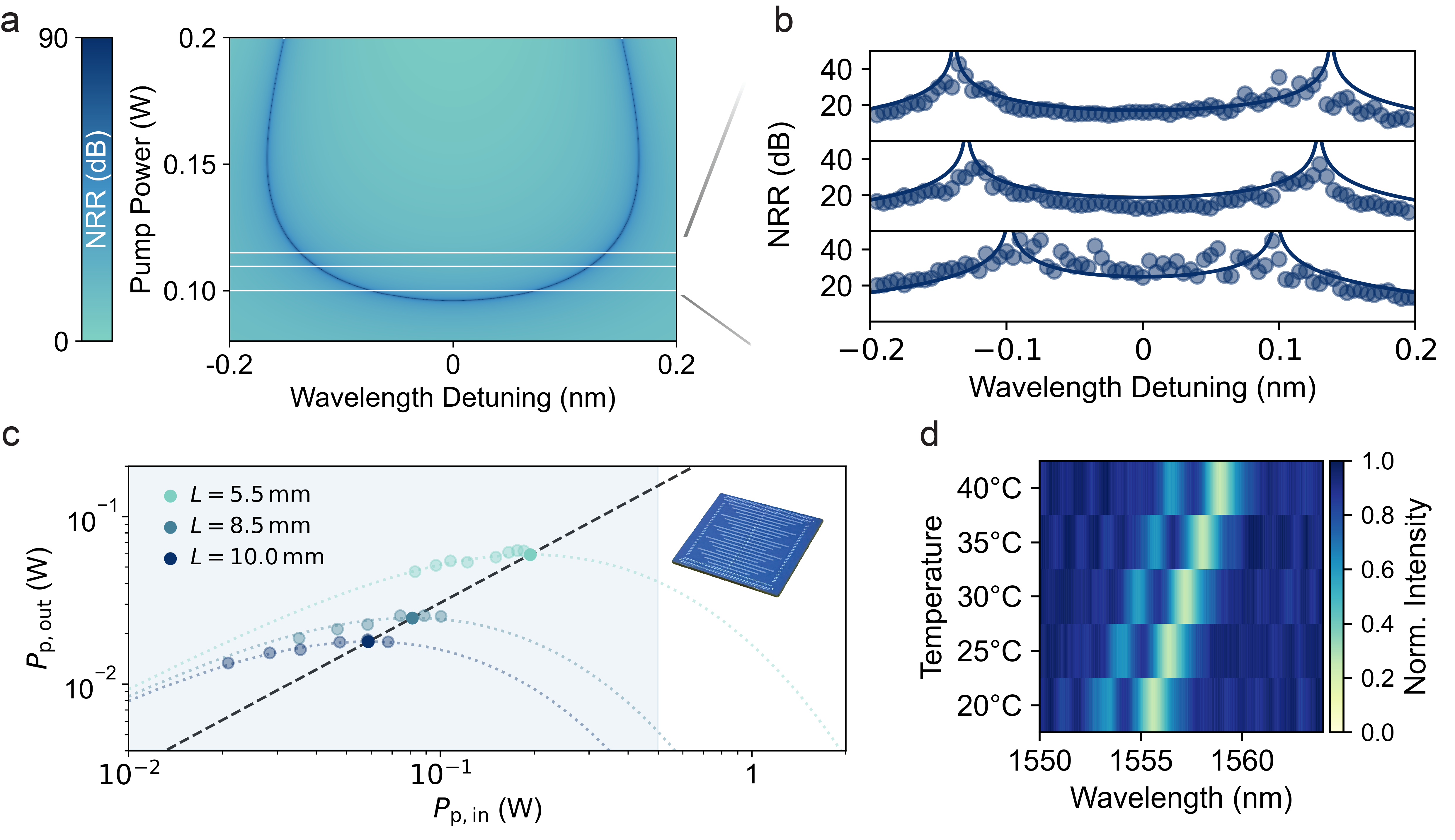}
\caption{{\bf Operation windows of PINE.} {\bf a}. A heatmap of simulated noise reduction ratio with respect to input pump power and pump wavelength detuning using experimentally derived $\eta$ and $L$ values. Three white lines indicate line slices at different input power levels. {\bf b}. Simulated (solid line) and measured (scatter points) NRR dependence on wavelength detuning from the QPM wavelength at three different pump powers (104 mW, 112 mW, 116 mW). The measured NRR are taken at 1 GHz RF frequency. {\bf c}. $P_{\text{p, in}}$ vs. $P_{\text{p, out}}$ dependence on different poling length on a single PINE chip. Faded scatter points indicate experimentally measured $P_{\text{p, out}}$ data. Inset showing the chip with different lengths of poling electrodes. The black dashed line indicates the critical power line that can be achieved by different poling lengths, and intersecting points with analytical $P_{\text{p, out}}$ solutions of the three different lengths are indicated by solid scatter points. {\bf d}. Temperature tuning of pump depletion at critical power, where the normalized FH intensity is $\approx 0.3$ at the QPM wavelength. The wavelength window shifts approximately linearly with the temperature of PINE chip.}
\label{fig:Fig4}
\end{figure*}
Intensity fluctuations are conveniently quantified by the relative-intensity-noise (RIN) spectrum, defined as the power spectral density of fractional power fluctuations, $S_{PP}(f)/P_0^2$, where $P_0$ is the average optical power. We quantify the device performance using the noise-reduction ratio
$\mathrm{NRR}(f) = 10\log_{10} [\mathrm{RIN}_{\mathrm{in}}(f)/\mathrm{RIN}_{\mathrm{out}}(f)]$, where positive values indicate suppression. In the small-signal limit, RIN at a Fourier frequency $f$ can be viewed as a small intensity modulation of the carrier at that frequency. We inject a calibrated sinusoidal intensity modulation with an electro-optic modulator driven by a vector network analyzer. By sweeping the modulation frequency from near DC to 10~GHz and comparing to light from the same laser that has not been sent through the PINE, we obtain $\mathrm{NRR}(f)$. The details of the measurement process and setup are shown in Methods and in Fig.~\ref{Sfig:VNA_Setup}.
The device exhibits robust passive stability, suppressing noise by $25$ to $60$~dB across the full measured spectrum (Fig.~\ref{fig:Fig3}). The apparent roll-off beyond $10$~GHz arises from detector and RF instrumentation limits rather than intrinsic device physics. To establish the fundamental bandwidth constraints, we developed a linearized sideband model treating intensity noise as weak modulation sidebands (Supplementary Information). This analysis reveals that the suppression bandwidth is governed by the group-velocity mismatch between the fundamental and second-harmonic waves: suppression degrades only when sidebands accrue significant dispersive phase slip relative to the carrier. The resulting 3-dB bandwidth scales as $f_{\mathrm{BW}} \sim  {1}/(2\pi L \left| 1/v_{g,s} - 1/v_{g,p} \right|)$. Thus, in contrast to active methods limited by electronic feedback bandwidth, PINE's bandwidth is set by the dispersive walk-off length, which can be engineered to support terahertz-scale operation as shown in Fig.~\ref{Sfig:GVM}.
\section{Operation Tolerance}
In addition to operating at the critical power and QPM wavelength, PINE supports a broader operational window. We explore this landscape by solving the SHG equations with wavelength detuning, generating a two-dimensional noise-suppression map (Fig.~\ref{fig:Fig4}a). This reveals that high noise suppression is not confined to a single point in the power-detuning space but extends over a range of conditions. At detunings away from phase matching, noise suppression is possible, albiet at higher powers, and accompanied by a conversion of intensity noise to phase noise which we study analytically in the supplementary materials (see Fig.~\ref{sfig:AMPMCoupling}). Fig.~\ref{fig:Fig4}b shows three power slices from this map overlaid with experimental data obtained by sweeping the pump wavelength. At higher input powers, the zero-noise transfer point splits into two spectral bands, allowing the operational wavelength bandwidth to be tuned simply by adjusting the input power. The experimental results show excellent agreement with simulation. For practical integration, the critical power at zero detuning can be engineered by varying the device length. We demonstrate this by fabricating PINE devices with different poling lengths (Fig.~\ref{fig:Fig4}c). These devices feature improved adaptive poling quality (Fig. \ref{Sfig:poling}), achieving a high normalized peak efficiency of $\eta \approx 2450 \% \si{W^{-1}cm^{-2}}$. The measured stationary points align well with analytical predictions (dashed line), covering a range compatible with on-chip lasers (blue shaded region). Furthermore, the operating wavelength can be actively shifted via temperature tuning (Fig.~\ref{fig:Fig4}d), enabling dynamic control of the stabilization window. 
\section{Shot Noise Limit}
\begin{figure}
\centering\includegraphics[width=\linewidth]{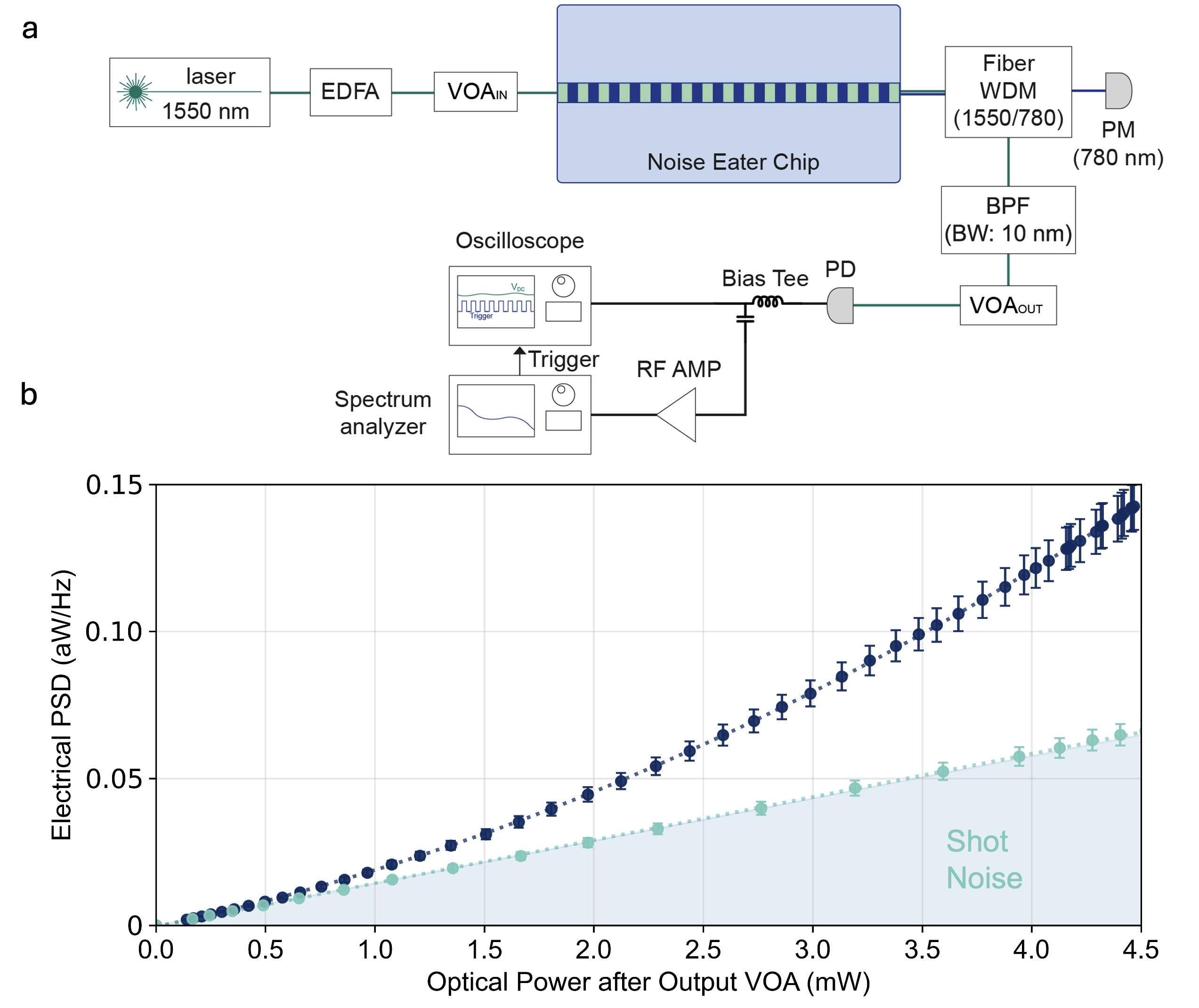}
\caption{{\bf Application of PINE approaching the quantum limit.} {\bf a}. A simplified schematic of the measurement set up for detecting shot noise from PINE. The setup is kept identical for measurements of EDFA and EDFA+PINE with an exception of bypassing PINE Chip (setup details shown in Fig. \ref{Sfig:Shot_Noise_Setup}). A fiber wavelength de-multiplexer (WDM) is used to separate FH and SH light. The FH output of WDM still has noticeable SH unfiltered. Therefore, a fiber bandpass filter (BPF) with 10 nm bandwidth is employed to filter the residual SH. Importantly, both measurements of EDFA only and EDFA with PINE see this BPF. Filtered FH power is connected to an output VOA. {\bf b}. Optical power dependence of PSD at the offset frequency of 100 MHz for EDFA and EDFA+PINE. The scatter points represent measured electrical PSD calibrated with averaged value of RF amplifier response over 1000 measurements, and the error bars represent the measured standard deviation. The navy scatter and fit line represent electrical PSD as a function of detected optical power for EDFA output alone, and the light green scatter and fit line represent corresponding data for EDFA output after PINE. The shaded region represents shot noise estimated at the given optical power.}
\label{fig:Fig5}
\end{figure}
Finally, we apply PINE to suppress the noise of an EDFA-amplified laser down to the standard quantum limit (SQL). This capability is critical for quantum sensing and precision metrology, where classical fluctuations often obscure quantum signals. Furthermore, deployable sensors in these fields require radically simplified optoelectronic noise suppression approaches compared to typical laboratory implementations. To verify this, we modified our setup to optimize collection efficiency and minimize the electronic noise floor (Fig.~\ref{fig:Fig5}a; see Methods). Key modifications included removing the noise-injection EOM, employing a fiber wavelength division multiplexer (WDM) to maximize collected power, and using a low-noise photodiode for direct power spectral density (PSD) measurements. We compared the PSD of the free-running EDFA output against the PINE-stabilized output at a 100 MHz offset frequency. By varying the detected power with a variable optical attenuator (VOA) \textit{after} the PINE device, we maintained the device at its critical operating point while characterizing the noise scaling. The free-running EDFA exhibits a quadratic dependence of PSD on optical power, characteristic of classical relative intensity noise (RIN). In stark contrast, the PINE-stabilized output follows a linear scaling (Fig.~\ref{fig:Fig5}b), the hallmark of shot-noise-limited behavior. This is further confirmed by the excellent agreement between our data and the theoretical shot-noise level (blue shaded region) calculated from the measured DC photocurrent. Looking forward, theoretical treatments of SHG suggest that such systems can evolve beyond the SQL to produce amplitude-squeezed states~\cite{kumar_li_1995_quantum_noise_evolution}. While current insertion losses limit our observation to the shot-noise floor, lower-loss implementations of PINE could unlock chip-scale sources of squeezed light for quantum information processing.

\section{Discussion}
We have demonstrated a second-harmonic-generation based nonlinear optical noise eater that achieves a broadband reduction in intensity-noise across flexible operating conditions. Utilizing highly efficient nonlinear frequency conversion in thin-film lithium niobate, the fundamental pump is depleted at relatively low optical powers, yielding power-stationary points at accessible optical power levels. The PINE thus provides a device-level solution for ultra-broadband intensity stabilization with a compact footprint ($ < 0.05$ mm $\times$ $10$ mm), operates without a resonator, and operates independent of specifics of the source laser. Noise reduction performance is competitive, providing a reduction $>$25 dB in $\text{RIN}$. Moreover, the reduction spans a broad RF bandwidth without active measurement and feedback, eliminating loop-bandwidth limits and the need for additional optoelectronic and signal processing components.

While the current performance is robust, several mechanisms likely limit the observed suppression to the 25-60 dB range. Back-reflections of the generated second-harmonic light could re-couple intensity fluctuations into the fundamental mode, degrading the noise performance. Furthermore, the passive nature of the device means that drifts in input coupling or laser power can shift the system away from the optimal critical power point. Future implementations could address these drifts using simple, low-bandwidth (sub-MHz) temperature or power stabilization loops. Additionally, advanced packaging techniques such as photonic wire bonding~\cite{lindenmann2012photonic} or heterogeneous integration~\cite{xiang2021laser, guo2022chip} would significantly improve coupling stability and reduce insertion losses.
Fundamentally, the intensity noise reduction is bounded by the laws of quantum mechanics. As the classical noise is suppressed, the system approaches the shot-noise limit, and potentially evolves into amplitude-squeezed states~\cite{kumar_li_1995_quantum_noise_evolution}. While our current measurements demonstrate operation at the quantum shot-noise limit, future work with lower-loss devices could explore these quantum regimes, potentially unlocking chip-scale sources of squeezed light.
In conclusion, PINE establishes a new practical, deployable approach for laser stabilization. By engineering the quasi-phase matching (shwon in Supplementary Information), we can tune the operating window to reduce insertion loss or broaden the wavelength acceptance. This technology paves the way for robust, quantum-limited local oscillators essential for next-generation atomic clocks, quantum sensors, and integrated photonic processors. 

\bibliography{Reference}

@article{bluvstein2024logical,
  title={Logical quantum processor based on reconfigurable atom arrays},
  author={Bluvstein, Dolev and Evered, Simon J and Geim, Alexandra A and Li, Sophie H and Zhou, Hengyun and Manovitz, Tom and Ebadi, Sepehr and Cain, Madelyn and Kalinowski, Marcin and Hangleiter, Dominik and others},
  journal={Nature},
  volume={626},
  number={7997},
  pages={58--65},
  year={2024},
  publisher={Nature Publishing Group UK London}
}

@article{atom_heating_noise,
  title = {Dynamics of noise-induced heating in atom traps},
  author = {Gehm, M. E. and O'Hara, K. M. and Savard, T. A. and Thomas, J. E.},
  journal = {Phys. Rev. A},
  volume = {58},
  issue = {5},
  pages = {3914--3921},
  numpages = {0},
  year = {1998},
  month = {Nov},
  publisher = {American Physical Society}
}

@article{atom_heating_noise_NIST_2025,
  title = {High Optical Access Cryogenic System for Rydberg Atom Arrays with a 3000-Second Trap Lifetime},
  author = {Zhang, Zhenpu and Hsu, Ting-Wei and Tan, Ting You and Slichter, Daniel H. and Kaufman, Adam M. and Marinelli, Matteo and Regal, Cindy A.},
  journal = {PRX Quantum},
  volume = {6},
  issue = {2},
  pages = {020337},
  numpages = {20},
  year = {2025},
  month = {May},
  publisher = {American Physical Society}
}

@article{LIGO_noise,
year = {2015},
month = {oct},
publisher = {IOP Publishing},
volume = {32},
number = {21},
pages = {215012},
author = {Powell, Jade and Trifirò, Daniele and Cuoco, Elena and Heng, Ik Siong and Cavaglià, Marco},
title = {Classification methods for noise transients in advanced gravitational-wave detectors},
journal = {Classical and Quantum Gravity},

}

@article{LO_Noise_Requirement,
  title = {Local-oscillator noise coupling in balanced homodyne readout for advanced gravitational wave detectors},
  author = {Steinlechner, Sebastian and Barr, Bryan W. and Bell, Angus S. and Danilishin, Stefan L. and Gl\"afke, Andreas and Gr\"af, Christian and Hennig, Jan-Simon and Houston, E. Alasdair and Huttner, Sabina H. and Leavey, Sean S. and Pascucci, Daniela and Sorazu, Borja and Spencer, Andrew and Strain, Kenneth A. and Wright, Jennifer and Hild, Stefan},
  journal = {Phys. Rev. D},
  volume = {92},
  issue = {7},
  pages = {072009},
  numpages = {8},
  year = {2015},
  month = {Oct},
  publisher = {American Physical Society}
}

@inbook{hall_taubman_ye_book,
  author = {John Hall and M. Taubman and Jun Ye},
  title = {Laser Stabilization},
  year = {2000},
  volume = {27},
  edition = {2},
  chapter = {27},
  number = {Handbook of Optics Series},
  pages = {27-1},
  month = {2000-01},
  publisher = {McGraw Hill Professional},
  issn = {0071364560},
  isbn = {9780071364560},
  note = {JILA Pub. 6435},
}

@inbook{active_noise_eater_book,
  title={Quantum noise: Basic measurements and techniques},
  author={Bachor, Hans A and Ralph, Timothy C},
  chapter = {8},
  pages = {269-301},
  year={2019},
  publisher={John Wiley \& Sons, Ltd},
}

@article{robertson1986intensity,
  title={Intensity stabilisation of an argon laser using an electro-optic modulator—Performance and limitations},
  author={Robertson, NA and Hoggan, S and Mangan, JB and Hough, J},
  journal={Applied Physics B},
  volume={39},
  number={3},
  pages={149--153},
  year={1986},
  publisher={Springer}
}

@article{xiang2021high,
  title={High-performance lasers for fully integrated silicon nitride photonics},
  author={Xiang, Chao and Guo, Joel and Jin, Warren and Wu, Lue and Peters, Jonathan and Xie, Weiqiang and Chang, Lin and Shen, Boqiang and Wang, Heming and Yang, Qi-Fan and others},
  journal={Nature communications},
  volume={12},
  number={1},
  pages={6650},
  year={2021},
  publisher={Nature Publishing Group UK London}
}

@article{nie2024turnkey,
  title={Turnkey photonic flywheel in a microresonator-filtered laser},
  author={Nie, Mingming and Musgrave, Jonathan and Jia, Kunpeng and Bartos, Jan and Zhu, Shining and Xie, Zhenda and Huang, Shu-Wei},
  journal={Nature Communications},
  volume={15},
  number={1},
  pages={55},
  year={2024},
  publisher={Nature Publishing Group UK London}
}

@article{lihachev2022low,
  title={Low-noise frequency-agile photonic integrated lasers for coherent ranging},
  author={Lihachev, Grigory and Riemensberger, Johann and Weng, Wenle and Liu, Junqiu and Tian, Hao and Siddharth, Anat and Snigirev, Viacheslav and Shadymov, Vladimir and Voloshin, Andrey and Wang, Rui Ning and others},
  journal={Nature communications},
  volume={13},
  number={1},
  pages={3522},
  year={2022},
  publisher={Nature Publishing Group UK London}
}

@article{casacio2021quantum_microscopy,
  title={Quantum-enhanced nonlinear microscopy},
  author={Casacio, Catxere A and Madsen, Lars S and Terrasson, Alex and Waleed, Muhammad and Barnscheidt, Kai and Hage, Boris and Taylor, Michael A and Bowen, Warwick P},
  journal={Nature},
  volume={594},
  number={7862},
  pages={201--206},
  year={2021},
  publisher={Nature Publishing Group UK London}
}

@article{siegman1962nonlinear,
  title={Nonlinear optical effects: an optical power limiter},
  author={Siegman, Anthony E},
  journal={Applied Optics},
  volume={1},
  number={S1},
  pages={127--132},
  year={1962},
  publisher={OSA}
}

@article{zia2025noise,
  title={Noise-immune quantum correlations of intense light},
  author={Zia Uddin, Shiekh and Rivera, Nicholas and Seyler, Devin and Sloan, Jamison and Salamin, Yannick and Roques-Carmes, Charles and Xu, Shutao and Sander, Michelle Y and Kaminer, Ido and Solja{\v{c}}i{\'c}, Marin},
  journal={Nature Photonics},
  pages={1--7},
  year={2025},
  publisher={Nature Publishing Group UK London}
}

@article{inoue2002experimental,
  title={Experimental study on noise characteristics of a gain-saturated fiber optical parametric amplifier},
  author={Inoue, Kyo and Mukai, Takaaki},
  journal={Journal of lightwave technology},
  volume={20},
  number={6},
  pages={969},
  year={2002},
  publisher={OSA}
}

@article{intensitynoise_optical_comm,
  title={Effect of intensity noise of semiconductor lasers on the digital modulation characteristics and the bit error rate of optical communication systems},
  author={Ahmed, Moustafa and Yamada, Minoru},
  journal={Journal of Applied Physics},
  volume={104},
  number={1},
  year={2008},
  publisher={AIP Publishing}
}

@article{RIN_in_optical_links,
  title = {Influence of Laser Relative-Intensity Noise on the Laser Interferometer Space Antenna},
  author = {Wissel, L. and Hartwig, O. and Bayle, J.B. and Staab, M. and Fitzsimons, E.D. and Hewitson, M. and Heinzel, G.},
  journal = {Phys. Rev. Appl.},
  volume = {20},
  issue = {1},
  pages = {014016},
  numpages = {15},
  year = {2023},
  month = {Jul},
  publisher = {American Physical Society}
}

@article{jankowski2024ultrafast,
  title={Ultrafast second-order nonlinear photonics—from classical physics to non-Gaussian quantum dynamics: a tutorial},
  author={Jankowski, Marc and Yanagimoto, Ryotatsu and Ng, Edwin and Hamerly, Ryan and McKenna, Timothy P and Mabuchi, Hideo and Fejer, MM},
  journal={Advances in Optics and Photonics},
  volume={16},
  number={2},
  pages={347--538},
  year={2024},
  publisher={Optica Publishing Group}
}

@article{SHG_1_5um_buffer,
  title={Intensity noise cancellation in solid-state laser at 1.5 $\mu$m using SHG depletion as a buffer reservoir},
  author={Audo, Kevin and Alouini, Mehdi},
  journal={Applied optics},
  volume={57},
  number={7},
  pages={1524--1529},
  year={2018},
  publisher={Optical Society of America}
}

@article{SHG_Buffer_Solid_State_Laser_2015,
author = {Abdelkrim El Amili and Mehdi Alouini},
journal = {Opt. Lett.},
number = {7},
pages = {1149--1152},
publisher = {Optica Publishing Group},
title = {Noise reduction in solid-state lasers using a SHG-based buffer reservoir},
volume = {40},
month = {Apr},
year = {2015}
}

@article{SHG_FH_Cavity_2024,
author = {Nanjing Jiao and Ruixin Li and Bingnan An and Jiawei Wang and Lirong Chen and Yajun Wang and Yaohui Zheng},
journal = {Opt. Lett.},
number = {13},
pages = {3568--3571},
publisher = {Optica Publishing Group},
title = {Passive laser power stabilization in a broadband noise spectrum via a second-harmonic generator},
volume = {49},
month = {Jul},
year = {2024}
}

@article{Porat_Freespace_Theory,
  title = {Broadband suppression of laser intensity noise based on second-harmonic generation},
  author = {Li, Y. and Seddighi, F. and Porat, G.},
  journal = {Phys. Rev. Appl.},
  volume = {22},
  issue = {1},
  pages = {014026},
  numpages = {11},
  year = {2024},
  month = {Jul},
  publisher = {American Physical Society}
}

@article{chen2024adaptivepoling,
  title={Adapted poling to break the nonlinear efficiency limit in nanophotonic lithium niobate waveguides},
  author={Chen, Pao-Kang and Briggs, Ian and Cui, Chaohan and Zhang, Liang and Shah, Manav and Fan, Linran},
  journal={Nature Nanotechnology},
  volume={19},
  number={1},
  pages={44--50},
  year={2024},
  publisher={Nature Publishing Group UK London}
}

@article{stokowski2024integrated,
  title={Integrated frequency-modulated optical parametric oscillator},
  author={Stokowski, Hubert S and Dean, Devin J and Hwang, Alexander Y and Park, Taewon and Celik, Oguz Tolga and McKenna, Timothy P and Jankowski, Marc and Langrock, Carsten and Ansari, Vahid and Fejer, Martin M and others},
  journal={Nature},
  volume={627},
  number={8002},
  pages={95--100},
  year={2024},
  publisher={Nature Publishing Group UK London}
}

@article{parameswaran2002observation,
  title={Observation of 99\% pump depletion in single-pass second-harmonic generation in a periodically poled lithium niobate waveguide},
  author={Parameswaran, Krishnan R and Kurz, Jonathan R and Roussev, Rostislav V and Fejer, Martin M},
  journal={Optics letters},
  volume={27},
  number={1},
  pages={43--45},
  year={2002},
  publisher={Optical Society of America}
}

@article{santandrea2019general_fidelity,
  title={General framework for the analysis of imperfections in nonlinear systems},
  author={Santandrea, Matteo and Stefszky, Michael and Silberhorn, Christine},
  journal={Optics Letters},
  volume={44},
  number={22},
  pages={5398--5401},
  year={2019},
  publisher={Optical Society of America}
}

@article{xiang2021laser,
  title={Laser soliton microcombs heterogeneously integrated on silicon},
  author={Xiang, Chao and Liu, Junqiu and Guo, Joel and Chang, Lin and Wang, Rui Ning and Weng, Wenle and Peters, Jonathan and Xie, Weiqiang and Zhang, Zeyu and Riemensberger, Johann and others},
  journal={Science},
  volume={373},
  number={6550},
  pages={99--103},
  year={2021},
  publisher={American Association for the Advancement of Science}
}

@article{guo2022chip,
  title={Chip-based laser with 1-hertz integrated linewidth},
  author={Guo, Joel and McLemore, Charles A and Xiang, Chao and Lee, Dahyeon and Wu, Lue and Jin, Warren and Kelleher, Megan and Jin, Naijun and Mason, David and Chang, Lin and others},
  journal={Science advances},
  volume={8},
  number={43},
  pages={eabp9006},
  year={2022},
  publisher={American Association for the Advancement of Science}
}

@article{lindenmann2012photonic,
  title={Photonic wire bonding: a novel concept for chip-scale interconnects},
  author={Lindenmann, N and Balthasar, G and Hillerkuss, D and Schmogrow, R and Jordan, M and Leuthold, Juerg and Freude, W and Koos, C},
  journal={Optics express},
  volume={20},
  number={16},
  pages={17667--17677},
  year={2012},
  publisher={Optical Society of America}
}

@article{PhysRev.127.1918,
  title = {Interactions between Light Waves in a Nonlinear Dielectric},
  author = {Armstrong, J. A. and Bloembergen, N. and Ducuing, J. and Pershan, P. S.},
  journal = {Phys. Rev.},
  volume = {127},
  issue = {6},
  pages = {1918--1939},
  numpages = {0},
  year = {1962},
  month = {Sep},
  publisher = {American Physical Society},
}

@article{boes2023lithium,
  title={Lithium niobate photonics: Unlocking the electromagnetic spectrum},
  author={Boes, Andreas and Chang, Lin and Langrock, Carsten and Yu, Mengjie and Zhang, Mian and Lin, Qiang and Lon{\v{c}}ar, Marko and Fejer, Martin and Bowers, John and Mitchell, Arnan},
  journal={Science},
  volume={379},
  number={6627},
  pages={eabj4396},
  year={2023},
  publisher={American Association for the Advancement of Science}
}

@article{park2024single,
  title={Single-mode squeezed-light generation and tomography with an integrated optical parametric oscillator},
  author={Park, Taewon and Stokowski, Hubert and Ansari, Vahid and Gyger, Samuel and Multani, Kevin KS and Celik, Oguz Tolga and Hwang, Alexander Y and Dean, Devin J and Mayor, Felix and McKenna, Timothy P and others},
  journal={Science Advances},
  volume={10},
  number={11},
  pages={eadl1814},
  year={2024},
  publisher={American Association for the Advancement of Science}
}

@article{wang2018integrated,
  title={Integrated lithium niobate electro-optic modulators operating at CMOS-compatible voltages},
  author={Wang, Cheng and Zhang, Mian and Chen, Xi and Bertrand, Maxime and Shams-Ansari, Amirhassan and Chandrasekhar, Sethumadhavan and Winzer, Peter and Lon{\v{c}}ar, Marko},
  journal={Nature},
  volume={562},
  number={7725},
  pages={101--104},
  year={2018},
  publisher={Nature Publishing Group UK London}
}

@article{santandrea2019general,
  title={General framework for the analysis of imperfections in nonlinear systems},
  author={Santandrea, Matteo and Stefszky, Michael and Silberhorn, Christine},
  journal={Optics Letters},
  volume={44},
  number={22},
  pages={5398--5401},
  year={2019},
  publisher={Optical Society of America}
}

@article{kumar_li_1995_quantum_noise_evolution,
  title={Evolution of quantum noise in the traveling-wave second-order [$\chi$ (2)] nonlinear process},
  author={Li, Ruo-Ding and Kumar, Prem},
  journal={Journal of the Optical Society of America B},
  volume={12},
  number={11},
  pages={2310--2320},
  year={1995},
  publisher={Optical Society of America}
}

\clearpage

\section{Methods}

\subsection*{Device Fabrication}
The geometry of the device (height and width of TFLN waveguides) and fabrication process of the devices are similar to those used in our prior work~\cite{stokowski2024integrated} with improved homogeneity of phase-matching by utilizing adaptive poling~\cite{chen2024adaptivepoling}. The detailed fabrication flow is shown in Fig.~\ref{Sfig:Fab_Flow}. On a thin-film of lithium niobate on insulator chip, we deposit a 100 nm protective layer of $\text{SiO}_{\text{2}}$ using high density plasma enhanced chemical vapor deposition. We then pattern Al electrodes, and use those electrodes to pole the LN with high voltage pulses. We then stripe Al electrodes and etch off the $\text{SiO}_{\text{2}}$. The waveguides are patterned using electron beam lithography, and etched using argon ion-mill dry etching. Finally, we clad the entire chip with $\text{SiO}_{\text{2}}$, and dice the facets for edge coupling. The waveguide geometry consists of a 1.2 $\mu \mathrm m$ ridge width and a 300 nm etch on a 500 nm film.

\subsection*{SHG Transfer function measurement}

\noindent
To measure the SHG transfer function, we use a setup similar to that described in reference ~\cite{stokowski2024integrated}, as shown in Fig.~\ref{fig:Fig5}a. We use a tunable C-band laser (Santec TSL-710) to couple pump power into the PINE chip via a lensed fiber. By scanning the laser wavelength across the QPM wavelength, we generate on-chip SH; both the generated SH and the residual FH are out-coupled from the chip via a lensed fiber. The collected light is then sent to a free-space dichroic mirror to split the FH and SH onto separate photodiodes, and we record both SH and FH power during the wavelength scan. The SHG fidelity is estimated following the procedure in~\cite{santandrea2019general_fidelity}, and a fit of the $\text{sinc}^2$ function on the SHG is shown in ~\ref{Sfig:CE_PD_Full} along with the FH transfer function for several different input powers, showing different levels of pump depletion.

Fidelity $\mathcal{F}$ is used to describe how close the transfer function is to an ideal $\text{sinc}^2$ function, and is defined as follows~\cite{santandrea2019general_fidelity}: 

\begin{equation}
    \mathcal{F}=\frac{\max \{V_{\text{SH}}/V_{\text{FH}}^2\}}{\int_{-\infty}^{\infty} V_{\text{SH}}/V_{\text{FH}}^2 \dd \lambda}\times\frac{2\pi}{|a|L}
\end{equation}

where $a=\frac{\partial\Delta k}{\partial\lambda}|_{\lambda=\lambda_{\text{QPM}}}$, $V_{\text{FH}}$ and $V_{\text{SH}}$ are the voltages from FH and SH photodetectors, respectively.

\subsection*{Noise Bandwidth Characterization}

We utilize a vector network analyzer (VNA) to inject intensity noise by driving an intensity modulator at the input. The detailed schematic of the measurement setup is shown in  ~\ref{Sfig:VNA_Setup}. Both input and output noise are measured with negligible delay and no change in the detection chain using a fiber MEMS switch. The VNA sweep uses 94 frequency points between 10 kHz and 10 GHz. At each point, a trigger is sent to the oscilloscope to accurately measure the detected DC voltage, which is then used to calibrate the RIN at that frequency. The maximum NRR exceeds 60 dB, and the NRR response remains high across the entire RF bandwidth (10 GHz).

\subsection*{Shot Noise Characterization}

\noindent
Shot-noise measurements are performed using the measurement setup shown in ~\ref{Sfig:Shot_Noise_Setup}. The laser output is amplified by an EDFA, which introduces RIN. The EDFA output is then coupled into PINE via a lensed fiber, and the PINE output is collected with a second lensed fiber. The collected optical power contains both SH and FH components; these are filtered using a custom Thorlabs fiber WDM (1550 nm / 780 nm) and a 10-nm-bandwidth bandpass filter (BPF) on the FH arm. The fiber WDM and BPF are used in fiber to improve the collection efficiency, avoiding a free-space dichroic and the associated coupling loss back into fiber. The FH power after the WDM + BPF is coupled into an output VOA, which provides controlled attenuation; we then record the PSD at different detected output powers.

An identical measurement is performed with the PINE chip bypassed by a single-mode fiber (SMF), as shown in the lower panel. The EDFA output in this bypass configuration is sent through the same detection chain (fiber WDM, BPF, VOA) to ensure consistency, and PSD data are again collected at different detected output powers.

We measure the electrical noise power spectral density (PSD) of the photodetected laser output using a spectrum analyzer with resolution bandwidth $\text{RBW} = $ 1 MHz. For each operating point, we record the DC voltage $V_{\text{DC}}$ across a load resistor $R_{\text{load}} = 50$ $\Omega$ on the oscilloscope after the bias tee. This voltage is used to infer the optical power incident on the photodiode,

\begin{equation}
    P_{\text{opt}} = \frac{V_{\text{DC}}}{R_{\text{load}} \mathcal{R}}
\end{equation}
where $\mathcal{R}$ is the detector responsivity, 0.9 A/W, which is the typical peak responsivity of the detector (Thorlab DET08CFC) used.
The measured SA noise level (in W, after conversion from dBm) is first
background-subtracted, and then referred back to the photodiode plane by dividing
out the total RF gain $G$ and normalizing by the SA resolution bandwidth:
\begin{equation}
    \text{PSD}_{\text{corr}} = \frac{\text{PSD}_{\text{bgsub}}}{G \times \text{RBW}}
\end{equation}
Here $\text{PSD}_{\text{bgsub}}$ is the measured electrical noise power with the background level subtracted, and $\text{PSD}_{\text{corr}}$ is the inferred electrical noise PSD (in W/Hz) referred to the photodiode. The corresponding uncertainty is obtained by propagating the measured gain
calibration error. If $\sigma_G$ is the standard deviation of the gain $G$
(in linear power units) of the RF amplifier, then
\begin{equation}
    \sigma_{\text{PSD}} = \text{PSD}_{\text{corr}} \cdot \frac{\sigma_G}{G}
\end{equation}
For comparison, we compute the fundamental shot-noise limit of the detector
chain. For optical power $P_{\text{opt}}$ on the photodiode, the average
photocurrent is
\begin{equation}
    I = \mathcal{R} P_{\text{opt}}
\end{equation}
and the (single-sided) shot-noise current PSD is
\begin{equation}
    S_I = 2 q I
\end{equation}
where $q$ is the elementary charge. This shot noise, when converted through the load resistor $R_{\text{load}}$, corresponds to an equivalent electrical power spectral density at the SA input,
\begin{equation}
    S_P(P_{\text{opt}}) = 2 q  \mathcal{R}  P_{\text{opt}}  R_{\text{load}}
\end{equation}
Since we start this estimation from the data set of (detected $V_{\text{DC}}$ and $\text{PSD}_{\text{corr}}$), we can accurately estimate the corresponding shot noise for the data set without having to know the responsivity. However, for presentation clarity, we use the estimated optical power for the plots, assuming the typical responsivity value of the photodiode. We plot $\text{PSD}_{\text{corr}}$ versus $P_{\text{opt}}$ for two cases: (i) EDFA only, and (ii) EDFA with PINE. We also plot
best-fit trends (quadratic for the EDFA case and linear for the EDFA+PINE case), and overlay the predicted shot-noise limit $S_P(P_{\text{opt}})$ as a shaded reference band.

\section*{Noise Reduction Ratio}

We write the optical power as
\begin{equation}
    P(t) = P_0 + \delta P(t)
\end{equation}
where $P_0$ is the average optical power and $\delta P(t)$ describes the intensity fluctuations.

We define $S_P(f)$ as the (single-sided) power spectral density (PSD) of the fluctuations $\delta P(t)$, with units of $\mathrm{W}^2/\mathrm{Hz}$. The frequency-resolved relative intensity noise (RIN) power spectral density is then
\begin{equation}
    S_{\mathrm{RIN}}(f) = \frac{S_P(f)}{P_0^2}
    \quad\left[\frac{1}{\mathrm{Hz}}\right]
\end{equation}

We measure the amplitude of this tone using a vector network analyzer (VNA). The measured $|S_{21}(f_0)|^2$ is proportional to the detected sideband power at $f_0$ after the common photodetector / transimpedance amplifier (TIA) / readout chain. Since we use an identical detection path, analyzer bandwidth, and source drive for two different device states $A$ and $B$ by using MEMS fiber switch, the ratio of the measured VNA responses directly tracks the ratio of the fluctuation amplitudes:
\begin{equation}
    \frac{\sigma_{P,\text{out}}}{\sigma_{P,\text{in}}}
    = \frac{|S_{21,\text{out}}(f_0)|}{|S_{21,\text{in}}(f_0)|}
\end{equation}

Combining this with the definition RIN$ = \sigma_P^2 / P_0^2$, we find that the ratio of the RIN between input power and output power is
\begin{equation}
    \text{NRR}=
    \frac{\text{RIN}_{\text{in}}}{\text{RIN}_{\text{out}}}
    =
    \frac{|S_{21,\text{out}}(f_0)|^2}{|S_{21,\text{in}}(f_0)|^2}
    \cdot
    \frac{P_{0,\text{in}}^2}{P_{0,\text{out}}^2}
    \label{eq:RIN_ratio}
\end{equation}
Equation~\eqref{eq:RIN_ratio} shows that, when the detection chain is kept identical, the relative RIN between two operating conditions can be obtained directly from the VNA power ratio and the ratio of the average detected electrical powers incident on the photodetector.
Generally speaking, the output power $P_{\text{out}}$ is a function of input power $P_{\text{in}}$ and wavelength detuning $\Delta\lambda$:
\begin{equation}
    P_{\text{out}}=f(P_{\text{in}},\Delta\lambda)
\end{equation}
For a fixed wavelength detuning $\Delta\lambda$, the output power fluctuation $\sigma_{P,\text{out}}$ can be approximated as:
\begin{equation}
    \sigma_{P,\text{out}}=\left|\frac{\partial f}{\partial P_{\text{in}}}\right|\sigma_{P,\text{in}}
\end{equation}
Thus we get:
\begin{equation}
\begin{aligned}
\text{NRR}_{\text{simulated}}
&=\left|\frac{\partial f}{\partial P_{\text{in}}}\right|^2 \cdot
  \left(\frac{P_{\text{in}}}{f(P_{\text{in}},\Delta\lambda)}\right)^2 \\
&=\left|\frac{\partial P_{\text{out}}}{\partial P_{\text{in}}}\right|^2 \cdot
  \left(\frac{P_{\text{in}}}{P_{\text{out}}}\right)^2
\end{aligned}
\end{equation}

\section*{Acknowledgments}
\noindent 

This work was supported by the U.S. Government via the Defense Advanced Research Projects Agency (DARPA) INSPIRED program (HR00112420356), and the National Science Foundation NSF-SNSF MOLINO project No. ECCS-2402483. Some of this work was funded by the US Department of Energy through grant no. DE-AC02-76SF00515 and via the Q-NEXT Center. Part of this work was performed at nano@stanford RRID:SCR\_026695. Part of this work was also performed at the Stanford Nano Shared Facilities (SNSF), supported by the National Science Foundation under award ECCS-2026822. D.D. acknowledges support from the NSF GRFP (No. DGE-1656518). H.S. acknowledges support from the Urbanek Family Fellowship. The authors are grateful for insightful discussions with Kevin Multani and Oguz Tolga Celik.

\section*{Author contributions}
G.H.A., Z.W., D.D., J.S., and A.S.N conceived of the project. G.H.A., Z.W. designed the device with inputs from D.D., and H.S.. Z.W. fabricated device with assistance from T.P.. G.H.A. and Z.W. performed experiments and analyzed data with inputs from D.D. and H.S.. M.M.F., J.S., and A.H.S.-N. provided experimental and theoretical support. A.H.S.-N. supervised the project. All authors contributed to data analysis and writing of the manuscript.

\section*{Competing interests}
\noindent 
G.H.A., Z.W., D.D., M.M.F., J.S.,and A.H.S.-N have filed a disclosure application (63/916962) for the nonlinear optical noise eater. The remaining authors declare no competing interests.

\section*{Additional information}
\noindent\textbf{Supplementary information} The online version contains supplementary material.

\noindent\textbf{Correspondence and requests for materials} should be addressed to Amir H. Safavi-Naeini.

\section*{Data availability}
\noindent Data will be made available upon request.

\clearpage
\onecolumngrid

\renewcommand{\figurename}{}
\renewcommand{\thefigure}{Extended Data Fig. \arabic{figure}}

\setcounter{figure}{0}

\begin{figure*}[p]
\centering\includegraphics[width=1\linewidth]{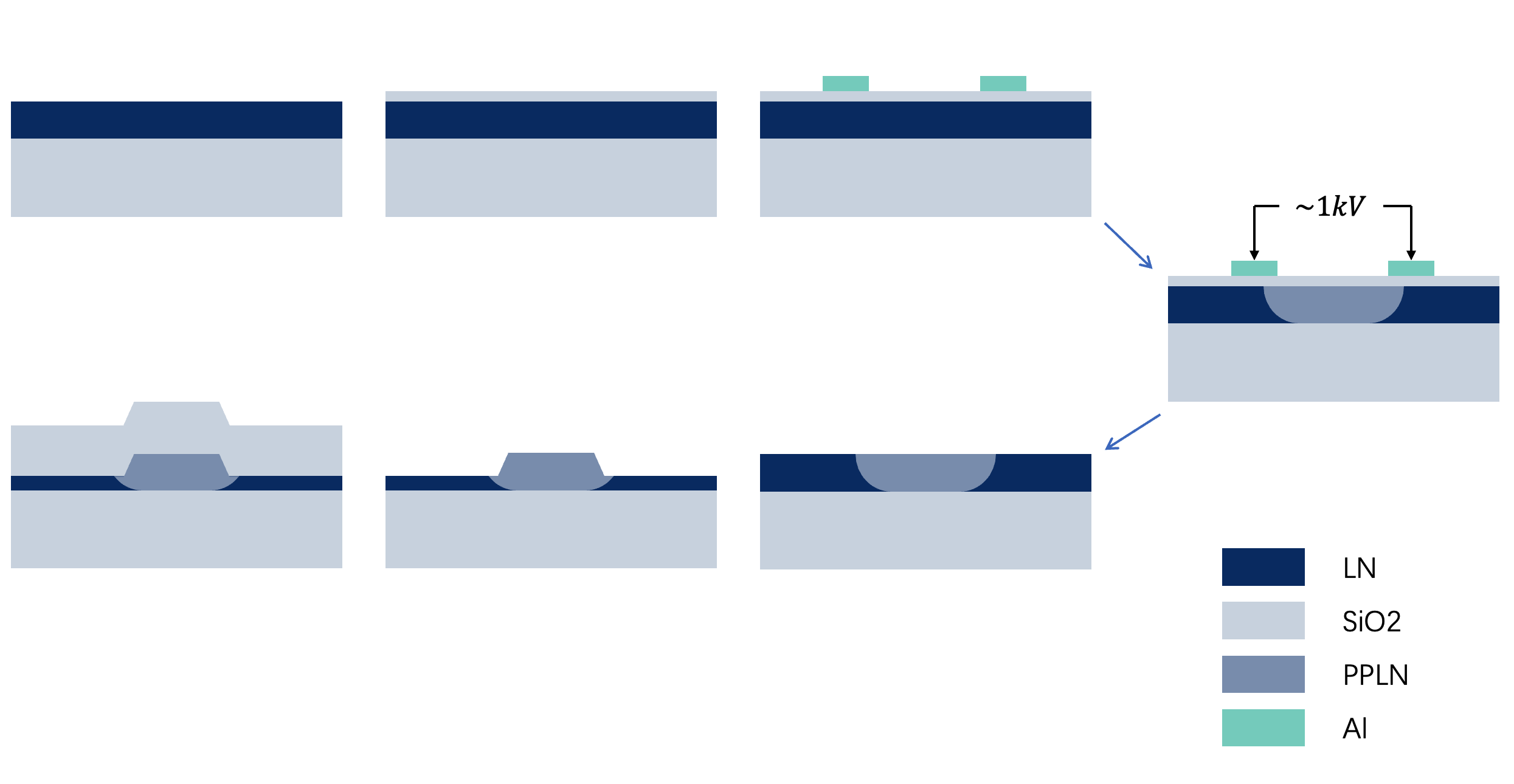}
\caption{{\bf Fabrication Process of PPTFLN integrated waveguides.} We start with TFLN sample, and silicon dioxide is deposited prior to poling electrodes deposition. Aluminium (Al) is used to pole the TFLN, and waveguide geometry is etched using argon ion-mill dry etching. The chip is then cladded with silicon dioxide.}
\label{Sfig:Fab_Flow}
\end{figure*}

\begin{figure*}[p]
\centering\includegraphics[width=0.7\linewidth]{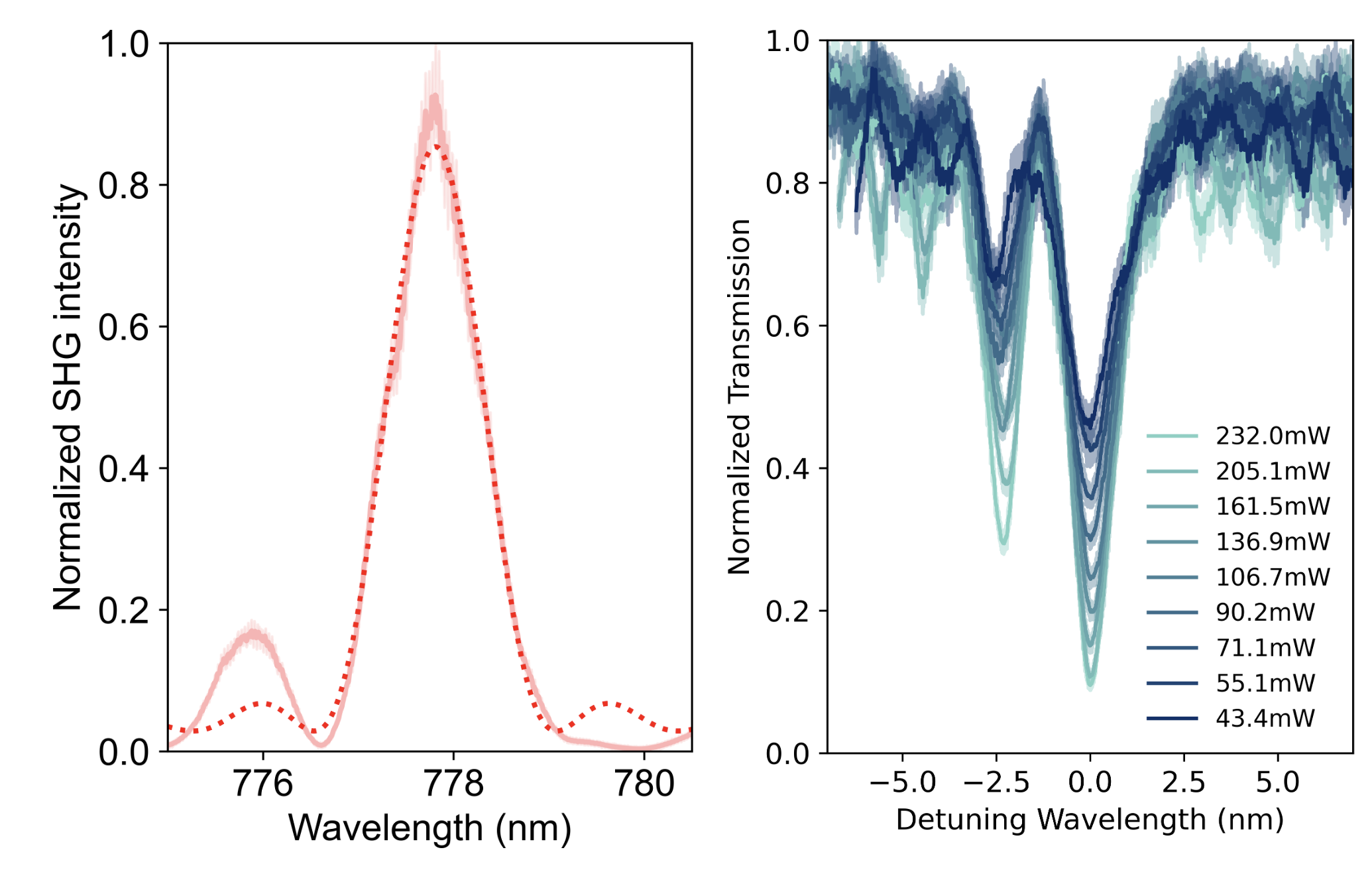}
\caption{{\bf Conversion Efficiency and Input Power-Dependent Pump Depletion}. {\bf a)} SHG intensity as a function of pump detuning; the x-axis displays the pump wavelength divided by 2. The peak efficiency is 1500 \% $\text{W}^{-1}\text{cm}^{-2}$. {\bf b)} Pump power dependent FH transfer function, showing greatly reduced normalized transmission at QPM wavelength due to pump depletion from SHG, where the transmission is normalized to maximum value across the wavelength scan.}
\label{Sfig:CE_PD_Full}
\end{figure*}

\begin{figure*}[p]
\centering\includegraphics[width=1\linewidth]{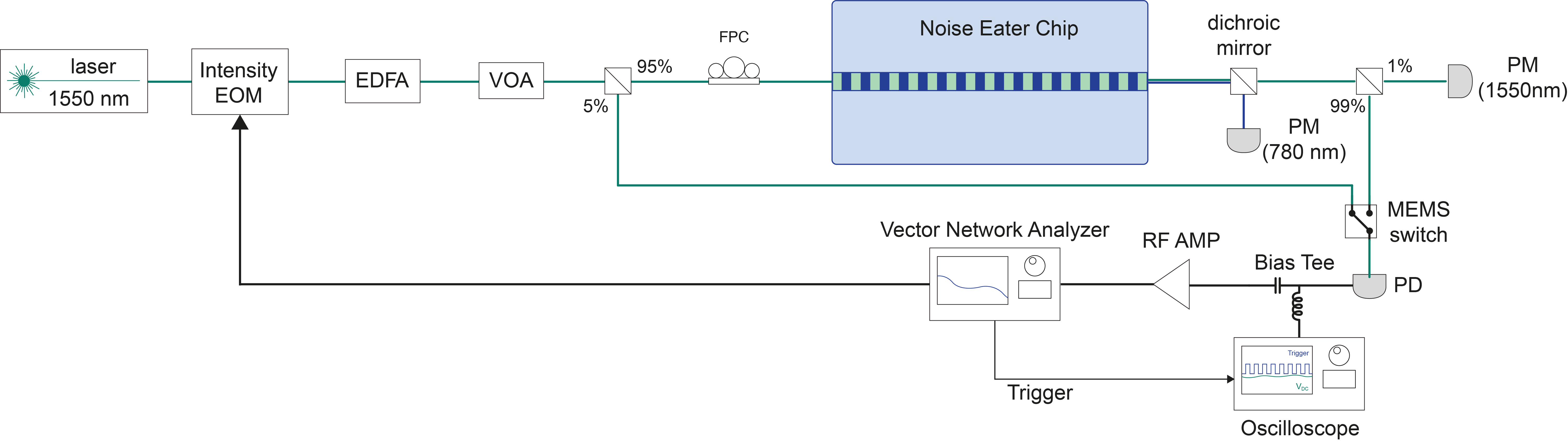}
\caption{{\bf Schematic of the Noise Bandwidth Characterization Setup.} Broadband noise reduction measurement set-up. EOM: Electro-Optic Modulator, EDFA: Erbium doped Fiber Amplifier, VOA: Variable Optical Attenuator, FPC: Fiber Polarization Controller, PM: Power Meter, PD: Photodiode, RF AMP: RF Amplifier. The 780 nm and 1550 nm signals are separately monitored, and utilized to find the CE $\approx 70\%$. A MEMS switch is utilized to switch between the input tap and output power of PINE. The VNA drives the intensity EOM at offset frequencies and triggers the oscilloscope to acquire precise DC voltage at a given VNA offset frequency point.}
\label{Sfig:VNA_Setup}
\end{figure*}

\begin{figure*}[p]
\centering\includegraphics[width=1\linewidth]{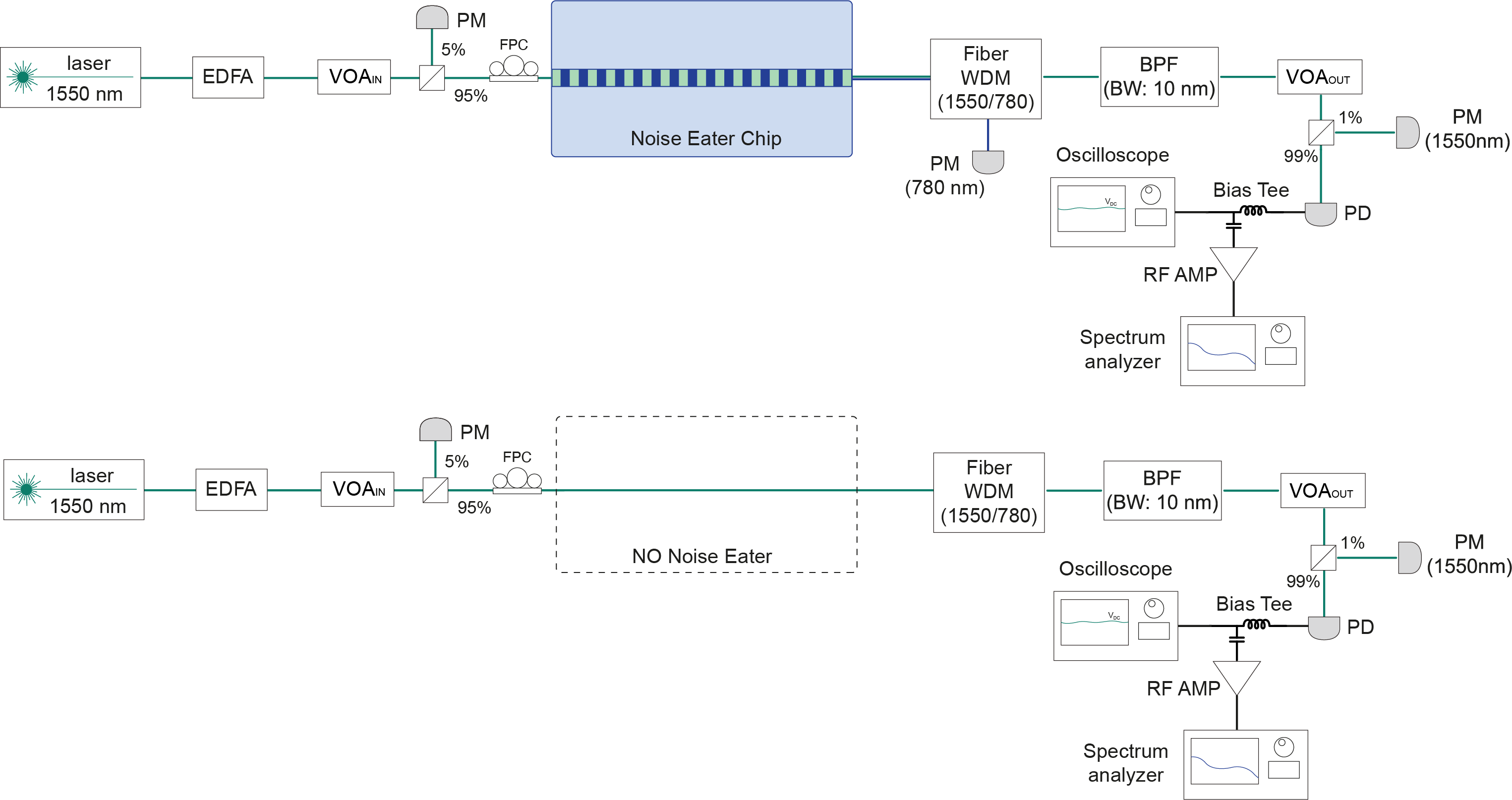}
\caption{{\bf Schematic of the Shot Noise Characterization Setup.} The top figure shows the full shot noise measurement set up with PINE chip, while the bottom figure shows the full shot noise measurement set up for input (EDFA only) calibration. }
\label{Sfig:Shot_Noise_Setup}
\end{figure*}

\clearpage
\onecolumngrid 

\appendix
\renewcommand{\figurename}{Fig.}
\renewcommand{\thefigure}{S\arabic{figure}}
\renewcommand{\thesection}{\Roman{section}}
\setcounter{figure}{0} 
\section* {Supplementary Information for \\ Ultrabroadband Passive Laser Noise Suppression to Quantum Noise Limit \\ through on-chip Second Harmonic Generation}
\vspace{-0.15 in}
\noindent Geun Ho Ahn$^{1, 2, \dagger} $, 
    Ziyu Wang$^{1, \dagger}$, 
    Devin J. Dean$^{1}$, 
    Hubert S. Stokowski$^{1}$,
    Taewon Park$^{1, 2}$, 
    Martin M. Fejer$^{1}$, 
    Jonathan Simon$^{1, 3}$,  
    Amir H. Safavi-Naeini$^{1, \ast}$\\
\vspace{0.1 in}\\
\noindent
$^1$Department of Applied Physics and Ginzton laboratory, Stanford University, Stanford, CA.\\
$^2$Department of Electrical Engineering, Stanford University, Stanford, CA.\\
$^2$Department of Physics, Stanford University, Stanford, CA.\\
{\small $^{\ast}$ safavi@stanford.edu
\section*{Contents}

\startcontents[post] 
\printcontents[post]{l}{1}[2]{}

\clearpage
\section*{Linearized Sideband Models of PINE}
\noindent
We start from the phase-matched, lossless three-wave \(\chi^{(2)}\) equations (no walk-off/GVD) for the fundamental-harmonic (FH) envelope \(A(z,t)\) at \(\omega_0\) and the second-harmonic (SH) envelope \(B(z,t)\) at \(2\omega_0\):
\begin{align}
\frac{\dd A}{\dd z} &= i \kappa A^{*} B, \label{eq:physA}\\
\frac{\dd B}{\dd z} &= i \kappa A^{2}. \label{eq:physB}
\end{align}
We write each field as a strong background (the SHG trajectory) plus small sidebands:
\begin{align}
A(z,t) &= A_0(z) + \delta A(z,t),\qquad 
B(z,t) = B_0(z) + \delta B(z,t), \label{eq:split}
\end{align}
where \(\delta A\) contains only the FH sidebands at \(\omega_0\pm\Omega\) and \(\delta B\) only the SH sidebands at \(2\omega_0\pm\Omega\). This assumption can be made as SH sidebands at \(2\omega_0\pm2\Omega\) are much smaller. 
\noindent
Thus we have:
\begin{equation}
    \delta A(z,t)=A_+(z)e^{-i\Omega t}+A_-(z)e^{i\Omega t},\quad \delta B(z,t)=B_+(z)e^{-i\Omega t}+B_-(z)e^{i\Omega t}
\end{equation}
Linearizing \eqref{eq:physB} about \( A_0\) gives
\begin{align}
\frac{\dd ( B_0+\delta B)}{\dd z} 
= i \kappa\big( A_0+\delta A\big)^2
= i \kappa\Big( A_0^2 + 2  A_0 \delta A + \mathcal O(\delta A^2)\Big).
\end{align}
The component at \(2\omega_0\pm\Omega\) originates only from the linear term \(2  A_0 \delta A\), hence at sideband level
\begin{equation}
\frac{\dd B_\pm}{\dd z} = i 2 \kappa  A_0 A_\pm\
\label{eq:railBz}
\end{equation}
Linearizing \eqref{eq:physA} yields, at \(\omega_0\pm\Omega\),
\begin{equation}
\frac{\dd A_\pm}{\dd z} = i \kappa\big( A_0^{*} B_\pm +  B_0 A_{\mp}^{*}\big)\; 
\label{eq:railAz}
\end{equation}
\noindent
Incorporating phase mismatch into Equations \ref{eq:physA}, \ref{eq:physB}, \ref{eq:railBz}, and \ref{eq:railAz}, we obtain:
\begin{equation}
    \begin{aligned}
        \frac{\dd A_+}{\dd z}&=i\kappa(A_0^*B_+e^{i\left(k_s(2\omega+\Omega)-k_p(\omega)-k_p(\omega+\Omega)-\frac{2\pi}{\Lambda}\right) z}+A_-^*B_0e^{i\left(k_s(2\omega)-k_p(\omega+\Omega)-k_p(\omega-\Omega)-\frac{2\pi}{\Lambda}\right) z})\\
        \frac{\dd A_-}{\dd z}&=i\kappa(A_0^*B_-e^{i\left(k_s(2\omega-\Omega)-k_p(\omega)-k_p(\omega-\Omega)-\frac{2\pi}{\Lambda}\right) z}+A_+^*B_0e^{i\left(k_s(2\omega)-k_p(\omega-\Omega)-k_p(\omega+\Omega)-\frac{2\pi}{\Lambda}\right) z})\\
        \frac{\dd B_+}{\dd z}&=2i\kappa A_0A_+e^{-i\left(k_s(2\omega+\Omega)-k_p(\omega)-k_p(\omega+\Omega)-\frac{2\pi}{\Lambda}\right) z}\\
        \frac{\dd B_-}{\dd z}&=2i\kappa A_0A_-e^{-i\left(k_s(2\omega-\Omega)-k_p(\omega)-k_p(\omega-\Omega)-\frac{2\pi}{\Lambda}\right) z}
    \end{aligned}
\end{equation}
and that:
\begin{equation}
    \begin{aligned}
        \frac{\dd A_0}{\dd z}&=i\kappa A_0^*B_0e^{i\left(k_s(2\omega)-2k_p(\omega)-\frac{2\pi}{\Lambda}\right) z}\\
        \frac{\dd B_0}{\dd z}&=i\kappa A_0^2e^{-i\left(k_s(2\omega)-2k_p(\omega)-\frac{2\pi}{\Lambda}\right) z}
    \end{aligned}
\end{equation}
We can make a Taylor expansion of the phase-mismatch terms:
\begin{equation}
    \begin{aligned}
        k_s(2\omega)-k_p(\omega+\Omega)-k_p(\omega-\Omega)-\frac{2\pi}{\Lambda}&\approx \Delta k -k_p^{\prime\prime}\Omega^2\\
        k_s(2\omega+\Omega)-k_p(\omega)-k_p(\omega+\Omega)-\frac{2\pi}{\Lambda}&\approx \Delta k +\left(k_s^\prime-k_p^\prime\right)\Omega+\frac{1}{2}\left(k_s^{\prime\prime}-k_p^{\prime\prime}\right)\Omega^2\\
        k_s(2\omega-\Omega)-k_p(\omega)-k_p(\omega-\Omega)-\frac{2\pi}{\Lambda}&\approx \Delta k -\left(k_s^\prime-k_p^\prime\right)\Omega+\frac{1}{2}\left(k_s^{\prime\prime}-k_p^{\prime\prime}\right)\Omega^2
    \end{aligned}
\end{equation}
where:
\begin{equation}
    \Delta k = k_s(2\omega)-2k_p(\omega)-\frac{2\pi}{\Lambda},\quad k_p^\prime=\frac{\dd k_p}{\dd \omega},\quad k_s^\prime=\frac{\dd k_s}{\dd \omega},\quad k_p^{\prime\prime}=\frac{\dd^2 k_p}{\dd \omega^2},\quad k_s^{\prime\prime}=\frac{\dd^2 k_s}{\dd \omega^2}
\end{equation}
Thus the equations become:
\begin{equation}
    \begin{aligned}
        \frac{\dd A_0}{\dd z}&=i\kappa A_0^*B_0e^{i\Delta k z}\\
        \frac{\dd B_0}{\dd z}&=i\kappa A_0^2e^{-i\Delta k z}\\
        \frac{\dd A_+}{\dd z}&=i\kappa(A_0^*B_+e^{i\left(\Delta k +\left(k_s^\prime-k_p^\prime\right)\Omega+\frac{1}{2}\left(k_s^{\prime\prime}-k_p^{\prime\prime}\right)\Omega^2\right) z}+A_-^*B_0e^{i\left(\Delta k -k_p^{\prime\prime}\Omega^2\right) z})\\
        \frac{\dd A_-}{\dd z}&=i\kappa(A_0^*B_-e^{i\left(\Delta k -\left(k_s^\prime-k_p^\prime\right)\Omega+\frac{1}{2}\left(k_s^{\prime\prime}-k_p^{\prime\prime}\right)\Omega^2\right) z}+A_+^*B_0e^{i\left(\Delta k -k_p^{\prime\prime}\Omega^2\right) z})\\
        \frac{\dd B_+}{\dd z}&=2i\kappa A_0A_+e^{-i\left(\Delta k +\left(k_s^\prime-k_p^\prime\right)\Omega+\frac{1}{2}\left(k_s^{\prime\prime}-k_p^{\prime\prime}\right)\Omega^2\right) z}\\
        \frac{\dd B_-}{\dd z}&=2i\kappa A_0A_-e^{-i\left(\Delta k -\left(k_s^\prime-k_p^\prime\right)\Omega+\frac{1}{2}\left(k_s^{\prime\prime}-k_p^{\prime\prime}\right)\Omega^2\right) z}
    \end{aligned}
\end{equation}
If we neglect the $\Omega^2$ terms, we obtain a simpler set of equations:
\begin{equation}
    \begin{aligned}
        \frac{\dd A_0}{\dd z}&=i\kappa A_0^*B_0e^{i\Delta k z}\\
        \frac{\dd B_0}{\dd z}&=i\kappa A_0^2e^{-i\Delta k z}\\
        \frac{\dd A_+}{\dd z}&=i\kappa(A_0^*B_+e^{i\left(\Delta k +\left(k_s^\prime-k_p^\prime\right)\Omega\right) z}+A_-^*B_0e^{i\Delta k z})\\
        \frac{\dd A_-}{\dd z}&=i\kappa(A_0^*B_-e^{i\left(\Delta k -\left(k_s^\prime-k_p^\prime\right)\Omega\right) z}+A_+^*B_0e^{i\Delta k z})\\
        \frac{\dd B_+}{\dd z}&=2i\kappa A_0A_+e^{-i\left(\Delta k +\left(k_s^\prime-k_p^\prime\right)\Omega\right) z}\\
        \frac{\dd B_-}{\dd z}&=2i\kappa A_0A_-e^{-i\left(\Delta k -\left(k_s^\prime-k_p^\prime\right)\Omega\right) z}
    \end{aligned}
\end{equation}

In the phase-matched case, i.e., $\Delta k=0$, we have:
\begin{equation}
    \begin{aligned}
        A_0(z)&=A_{p,\text{in}}\sech{(\kappa A_{p,\text{in}} z)}\\
        B_0(z)&=iA_{p,\text{in}}\tanh{(\kappa A_{p,\text{in}} z)}
    \end{aligned}
\end{equation}
And the equations for 1-st order sidebands become:
\begin{equation}
    \begin{aligned}
        \frac{\dd A_+}{\dd z}&=i\kappa\left(A_{p,\text{in}}\sech{(\kappa A_{p,\text{in}} z)}\right)^*B_+e^{i\left(k_s^\prime-k_p^\prime\right)\Omega z}+i\kappa A_-^*\left(iA_{p,\text{in}}\tanh{(\kappa A_{p,\text{in}} z)}\right)\\
        \frac{\dd A_-}{\dd z}&=i\kappa\left(A_{p,\text{in}}\sech{(\kappa A_{p,\text{in}} z)}\right)^*B_-e^{-i\left(k_s^\prime-k_p^\prime\right)\Omega z}+i\kappa A_+^*\left(iA_{p,\text{in}}\tanh{(\kappa A_{p,\text{in}} z)}\right)\\
        \frac{\dd B_+}{\dd z}&=2i\kappa \left(A_{p,\text{in}}\sech{(\kappa A_{p,\text{in}} z)}\right)A_+e^{-i\left(k_s^\prime-k_p^\prime\right)\Omega z}\\
        \frac{\dd B_-}{\dd z}&=2i\kappa \left(A_{p,\text{in}}\sech{(\kappa A_{p,\text{in}} z)}\right)A_-e^{i\left(k_s^\prime-k_p^\prime\right)\Omega z}
    \end{aligned}
\end{equation}
For simplicity, we can define
\begin{equation}
\label{eq:ab_definitions}
    a_{\pm}=\frac{A_\pm}{A_{p,\text{in}}},\quad b_{\pm}=\frac{B_\pm}{A_{p,\text{in}}},\quad \zeta=\kappa A_{p,\text{in}} z\in\left[0,\kappa A_{p,\text{in}}L\right],\quad k^\prime=\frac{k_s^\prime-k_p^\prime}{\kappa A_{p,\text{in}}}\Omega
\end{equation}
Then we get:
\begin{equation}
    \label{equ:linear_perturbation}
    \begin{aligned}
        \frac{\dd a_+}{\dd \zeta}&=i\sech{(\zeta)}b_+e^{ik^\prime \zeta}-a_-^*\tanh{(\zeta)}\\
        \frac{\dd a_-}{\dd \zeta}&=i\sech{(\zeta)}b_-e^{-ik^\prime \zeta}-a_+^*\tanh{(\zeta)}\\
        \frac{\dd b_+}{\dd \zeta}&=2ia_+\sech{(\zeta)}e^{-ik^\prime \zeta}\\
        \frac{\dd b_-}{\dd \zeta}&=2ia_-\sech{(\zeta)}e^{ik^\prime \zeta}
    \end{aligned}
\end{equation}

With these linearized ODEs, we can numerically solve for $|a_{\pm}|$ which represent noise away from the carrier, and is plotted as a function of $\zeta$ in Fig. \ref{Sfig:sideband_zeta}.
\begin{figure}[h!]
\centering\includegraphics[width=0.9\linewidth]{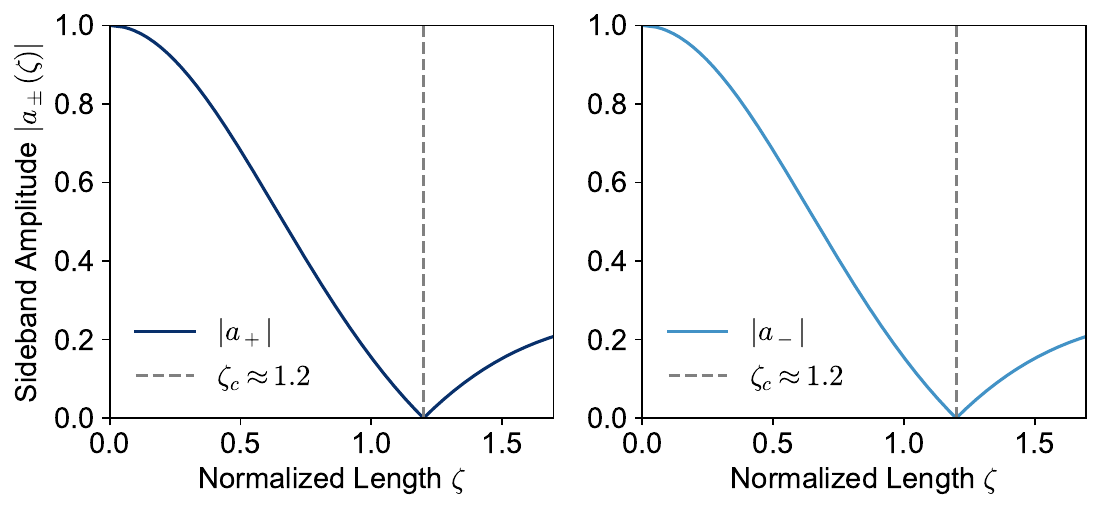}
\caption{{\bf Simulated amplitude of FH sidebands using the linearized sideband model.}}
\label{Sfig:sideband_zeta}
\end{figure}

This shows that at $\zeta=\zeta_c \approx 1.2$, the sideband amplitudes are completely suppressed, and this is consistent with our DC analysis done with time-domain simulation of SHG in the Fig.~\ref{fig:Fig1}b, and Equation (3). 

In the presence of both amplitude-modulation and phase-modulation, the input field is:
\begin{equation}
    \begin{aligned}
    \label{eq:input_field_expression1}
        A(0,t)&=\sqrt{P_{p,\text{in}}}\left(1+\varepsilon_{\text{AM,in}}\cos\Omega t\right)e^{i\varepsilon_{\text{PM,in}}\cos(\Omega t+\phi_{\text{in}})}\\
        &\approx \sqrt{P_{p,\text{in}}}\left(1+\varepsilon_{\text{AM,in}}\cos\Omega t+i\varepsilon_{\text{PM,in}}\cos(\Omega t+\phi_{\text{in}})\right)\\
        &=\sqrt{P_{p,\text{in}}}\left[1+\frac{1}{2}\left(\varepsilon_{\text{AM,in}}+i\varepsilon_{\text{PM,in}}e^{-i\phi_{\text{in}}}\right)e^{-i\Omega t}+\frac{1}{2}\left(\varepsilon_{\text{AM,in}}+i\varepsilon_{\text{PM,in}}e^{i\phi_{\text{in}}}\right)e^{i\Omega t}\right]
    \end{aligned}
\end{equation}
Writing the input field as:
\begin{equation}
\label{eq:input_field_expression2}
    A(0,t)=A_{p,\text{in}}\left(1+a_+(0)e^{-i\Omega t}+a_-(0)e^{i\Omega t}\right)
\end{equation}
and comparing Equation \ref{eq:input_field_expression1} and Equation \ref{eq:input_field_expression2}, we find:
\begin{equation}
\label{eq:a_pm0}
   A_{p,\text{in}}=\sqrt{P_{p,\text{in}}},\quad a_\pm(0)=\frac{1}{2}\left(\varepsilon_{\text{AM,in}}+i\varepsilon_{\text{PM,in}}e^{\mp i\phi}\right)
\end{equation}
Using $a_{\pm}(0)$ to represent $\varepsilon_{\text{AM,in}}$, $\varepsilon_{\text{PM,in}}$ and $\phi_{\text{in}}$, we get:
\begin{equation}
\label{equ:from_a_to_epsilon}
    \begin{aligned}
        \varepsilon_{\text{AM,in}}&=\text{Re}{\{a_+(0)+a_-(0)\}}\\
        \varepsilon_{\text{PM,in}}&=\sqrt{\text{Im}\{a_+(0)+a_-(0)\}^2+\text{Re}{\{a_+(0)-a_-(0)\}}^2}\\
        \phi_{\text{in}}&=\arctan{\left(\frac{\text{Re}{\{a_+(0)-a_-(0)\}}}{\text{Im}\{a_+(0)+a_-(0)\}}\right)}
    \end{aligned}
\end{equation}
With $a_\pm (0)$ as given in terms of AM and PM amplitudes in Equation \ref{eq:a_pm0}, and $b_\pm(0)=0$, we can solve Equation \ref{equ:linear_perturbation} numerically for $a_\pm (L)$. With Equation \ref{eq:ab_definitions}, the output sidebands are:
\begin{equation}
    A_\pm(L)=A_{p,\text{in}}a_\pm (L)=\sqrt{P_{p,\text{in}}}a_\pm (L)
\end{equation}

Recalling Equation \ref{eq:input_field_expression2} and that $A_0(L)=A_{p,\text{in}}\sech\left(\kappa A_{p,\text{in}}L\right)$, the total output field is:
\begin{equation}
    A(L,t)=A_0(L)\left(1+\cosh{\left(\kappa A_{p,\text{in}}L\right)}\left[a_+(L)e^{-i\Omega t}+a_-(L)e^{i\Omega t}\right]\right)
\end{equation}
We can rewrite this field in terms of its in-phase and quadrature parts according to:
\begin{equation}
    A(L,t)=A_{p,\text{out}}\left[1+\varepsilon_{\text{AM,out}}\cos\left(\Omega t\right)+i\varepsilon_{\text{PM,out}}\cos\left(\Omega t+\phi_{\text{out}}\right)\right]
\end{equation}
where with some algebra we find:
\begin{equation}
    \begin{aligned}
        \varepsilon_{\text{AM,out}}&=\text{Re}{\{a_+(L)+a_-(L)\}}\cosh{\left(\kappa A_{p,\text{in}}L\right)}\\
        \varepsilon_{\text{PM,out}}&=\sqrt{\text{Im}\{a_+(L)+a_-(L)\}^2+\text{Re}{\{a_+(L)-a_-(L)\}}^2}\cosh{\left(\kappa A_{p,\text{in}}L\right)}\\
        \phi_{\text{out}}&=\arctan{\left(\frac{\text{Re}{\{a_+(L)-a_-(L)\}}}{\text{Im}\{a_+(L)+a_-(L)\}}\right)}
    \end{aligned}
\end{equation}
Thus the frequency-resolved transferred RIN could be written as:
\begin{equation}
    \text{RIN}_{\text{transfer}}=\left|\frac{\Delta P_{\text{out}}/P_{\text{out}}}{\Delta P_{\text{in}}/P_{\text{in}}}\right|^2=\left|\frac{\varepsilon_{\text{AM,out}}}{\varepsilon_{\text{AM,in}}}\right|^2
\end{equation}
and the frequency-resolved transferred PM amplification factor could be written as:
\begin{equation}
    \text{PM}_{\text{transfer}}=\left|\frac{\varepsilon_{\text{PM,out}}}{\varepsilon_{\text{PM,in}}}\right|^2
\end{equation}
Intensity noise reduction and phase noise reduction are defined as
\begin{equation}
    \begin{aligned}
        \text{Intensity Noise Reduction (dB)}&= -10 \log_{10}\left(\text{RIN}_{\text{transfer}}\right)\\
        \text{Phase Noise Reduction (dB)}&= -10 \log_{10}\left(\text{PM}_{\text{transfer}}\right)\\
    \end{aligned}
\end{equation}
\noindent We also solve the ODEs numerically to get RF transfer functions for both AM and PM cases.
Setting $\varepsilon_{\text{AM,in}}=\varepsilon_{\text{PM,in}}=10^{-3}$, we obtain the noise reduction over the RF bandwidth shown in Fig.~\ref{Sfig:SI_RF_Bandwidth}. The degradation in noise reduction ratio arises from the phase slip $k^\prime$, which depends on the group velocity mismatch. Fig.~\ref{Sfig:GVM} shows that by reducing the group velocity mismatch values, we can reach THz noise reduction bandwidth.

\begin{figure}
    \centering
    \includegraphics[width=\textwidth]{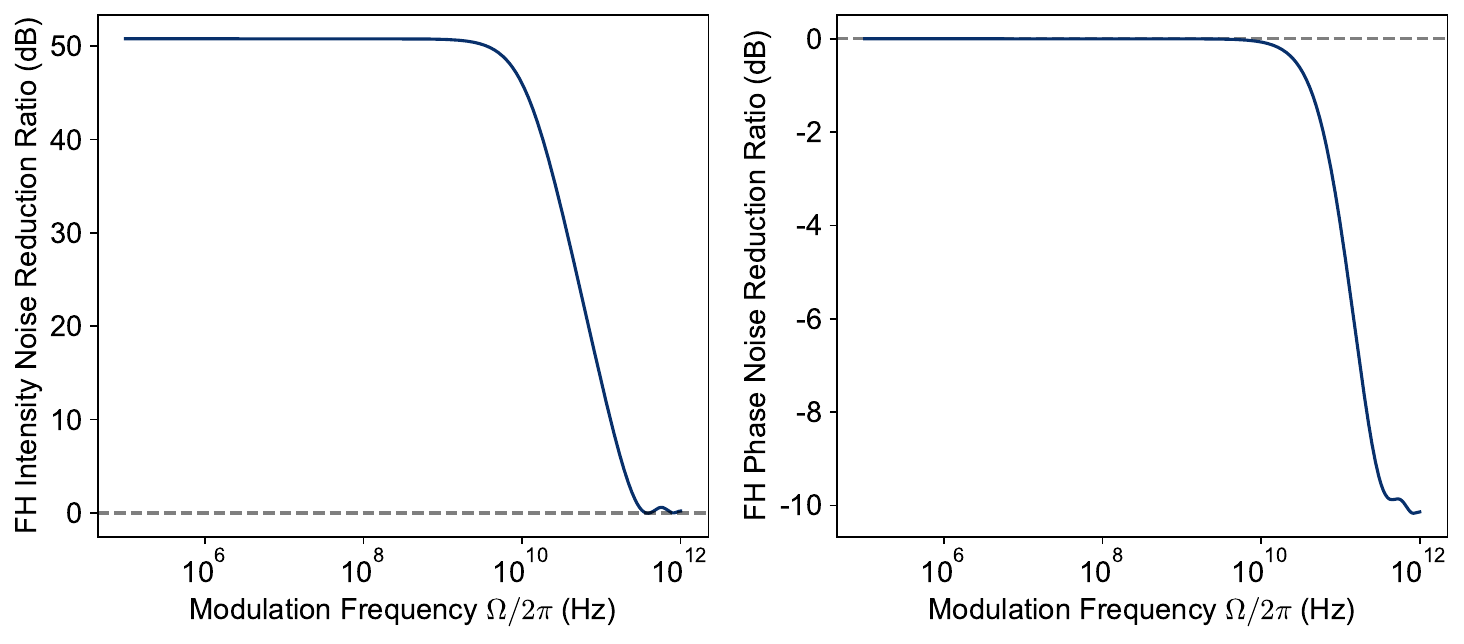}
    \caption{RF transfer functions for both AM and PM cases, calculated by numerically solving the ODEs (Equation \ref{equ:linear_perturbation}). Other parameters are: $P_{p,\text{in}}=99.6\% \times P_{p,\text{c}}$, $L=10\si{mm}$, $\eta=\kappa^2=1500\%{\text{W}^{-1}\text{cm}^{-2}}$, $n_{g,p}=2.249681$, $n_{g,s}=2.322913$, $k^\prime_{p/s}=n_{g,p/s}/c_0$.}
    \label{Sfig:SI_RF_Bandwidth}
\end{figure}

\begin{figure}
    \centering
    \includegraphics[width=0.8\textwidth]{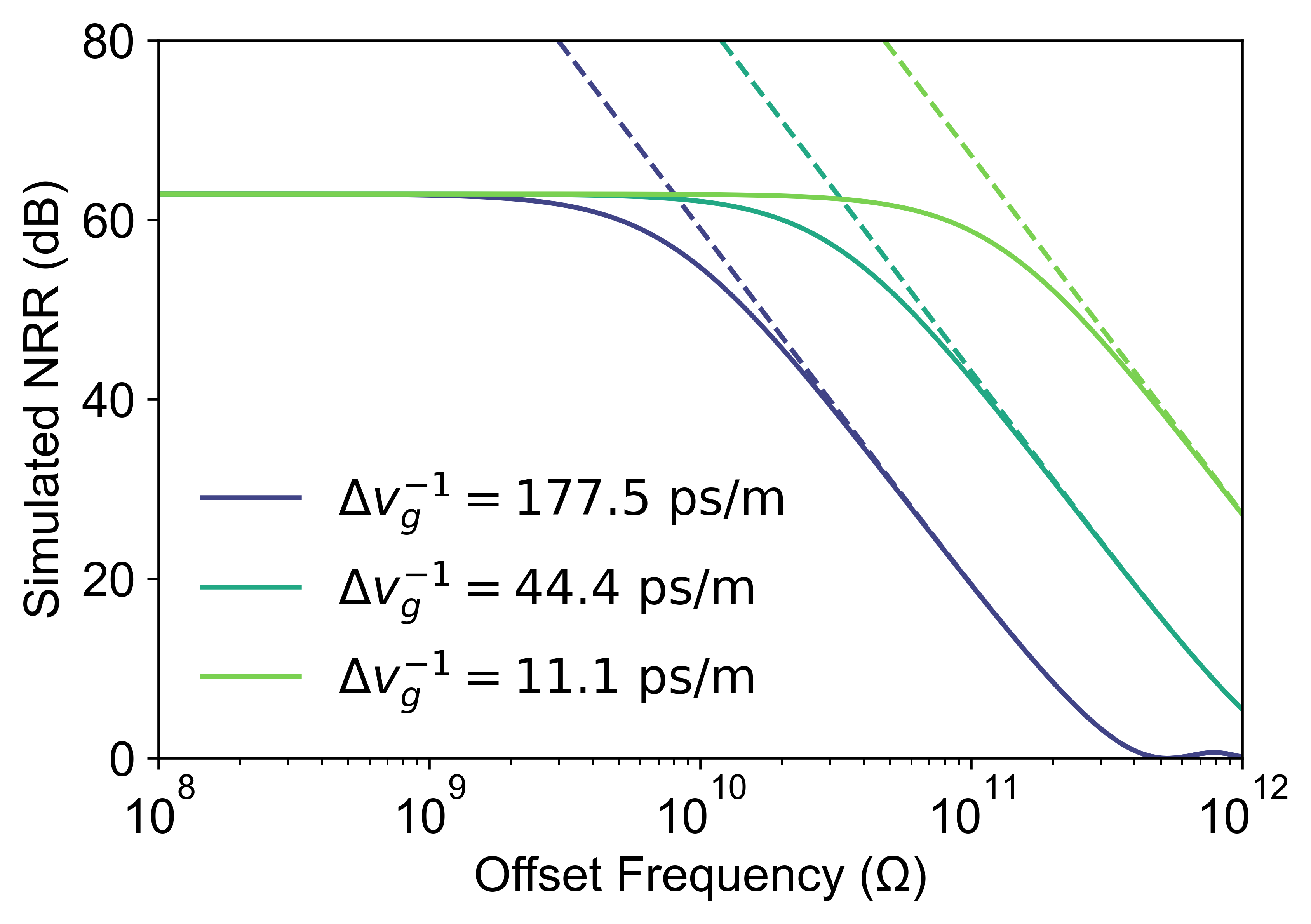}
    \caption{GVM dependence of RF transfer functions for AM, calculated by numerically solving the ODEs (Equation. \ref{equ:linear_perturbation}) with different group velocity mismatch values. Solid lines depict noise eater operation at the optical power of 99.9$\% \times P_{p,\text{c}}$, while dotted lines indicate at critical power.}
    \label{Sfig:GVM}
\end{figure}
AM noise reduction can be studied with different power offsets, as shown below. 

\begin{figure}[h!]
\centering
\includegraphics[width=\linewidth]{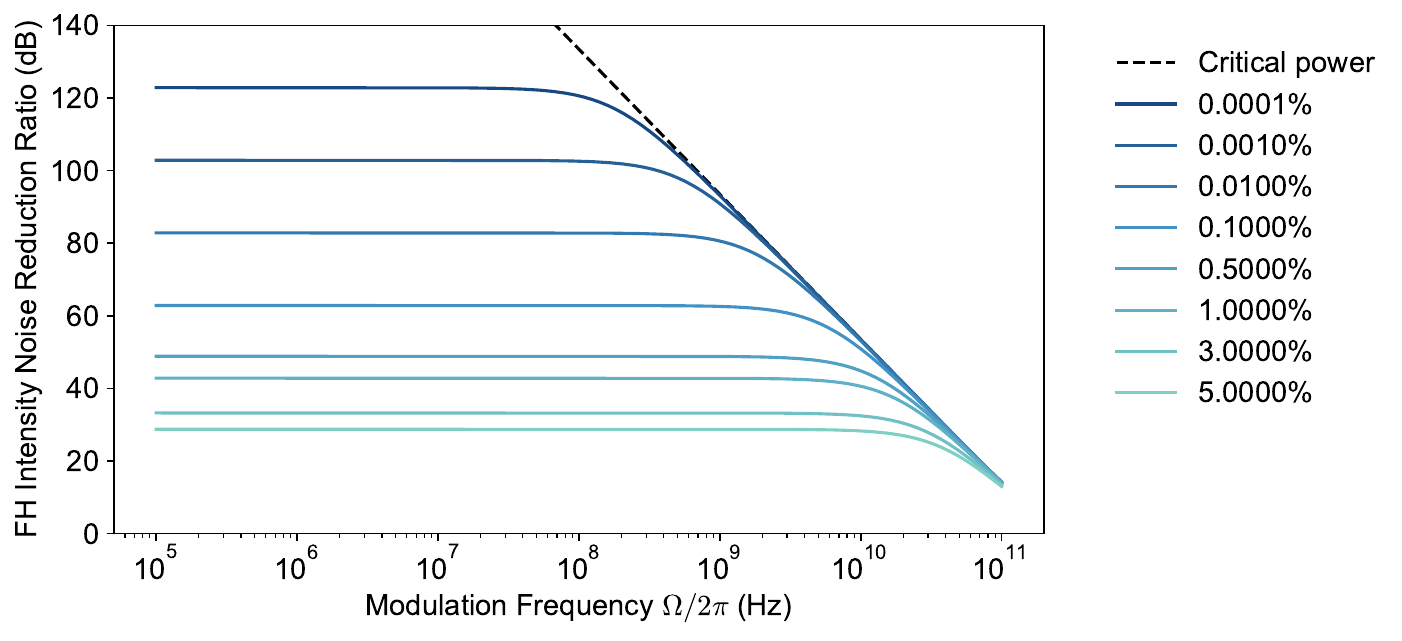}
\caption{{\bf Simulated intensity NRR at different power offsets ranging from 0.0001\% to the critical power.}}
\label{Sfig:power_fraction}
\end{figure}

\newpage
\newpage
\clearpage

\section*{Analytical Solutions to Nonlinear Wave Equations for Wavelength and Power Operation Window of PINE}

Again, the envelop equations for FH and SH are as follows:
\begin{equation}
    \begin{aligned}
        \frac{\dd A}{\dd z}&=i\kappa A^*B e^{i\Delta k z}\\
        \frac{\dd B}{\dd z}&=i\kappa A^2 e^{-i\Delta k z}\\
        \Delta k&=k_B-2k_A-2\pi/\Lambda=a\left(\lambda-\lambda_{\text{QPM}}\right)
    \end{aligned}
\end{equation}
with initial conditions:
\begin{equation}
    \begin{aligned}
        A(0)=\sqrt{P_{p,\text{in}}}e^{i\theta_p(0)},\quad B(0)=0
    \end{aligned}
\end{equation}
We can write the variables as:
\begin{equation}
    A(z)=|A(z)|e^{i\theta_p(z)},\quad B(z)=|B(z)|e^{i\theta_s(z)}
\end{equation}

\noindent With the following transposition:
\begin{equation}
    u(z)=\kappa |A(z)|l,\  v(z)=\kappa |B(z)|l,\  \theta(z)=\theta_s(z)-2\theta_p(z)+\Delta kz, \  l=\frac{1}{\sqrt{\kappa^2 P_{p,\text{in}}}},\  \zeta=\frac{z}{l}
\end{equation}

We can the following equations describing the evolution of $u,v,\theta_p,\theta_s$ and $\theta$:
\begin{equation}
    \label{equ:dimensionless-equations}
    \begin{aligned}
        \frac{\dd u}{\dd \zeta}&=-uv\sin\theta,\quad\frac{\dd \theta_p}{\dd \zeta}=v\cos\theta\\
        \frac{\dd v}{\dd \zeta}&=u^2\sin\theta,\quad\frac{\dd \theta_s}{\dd \zeta}=\frac{u^2}{v}\cos\theta\\
        \frac{\dd \theta}{\dd \zeta}&=\Delta s+\frac{\cos\theta}{\sin\theta}\frac{\dd }{\dd \zeta}\ln{\left(u^2v\right)}
    \end{aligned}
\end{equation}

By solving Equation. \ref{equ:dimensionless-equations}, we can obtain the output power and phase for both FH and SH, in the form of Jacobi elliptic functions~\cite{PhysRev.127.1918}:

\begin{equation}
    \label{equ:analytical_sol}
    \begin{aligned}
        P_{p,\text{out}}&=|A(L)|^2=\frac{u^2(\zeta_L)}{\kappa^2l^2}=u^2(\zeta_L) P_{p,\text{in}}=\left(1-v_b^2\text{sn}^2(v_c\zeta_L,\gamma^2)\right)P_{p,\text{in}}\\
        P_{s,\text{out}}&=|B(L)|^2=\frac{v^2(\zeta_L)}{\kappa^2l^2}=v^2(\zeta_L) P_{p,\text{in}}=v_b^2\text{sn}^2(v_c\zeta_L,\gamma^2)P_{p,\text{in}}\\
        \theta_s(L)&=\frac{\pi}{2}-\frac{1}{2}\Delta kL+2\theta_p(0)\\
        \theta_p(L)&=\frac{1}{4}\Delta kL-\frac{1}{2}\arcsin\left(\frac{\Delta kl}{2}\frac{v(\zeta_L)}{u^2(\zeta_L)}\right)+\theta_p(0)\\
    \end{aligned}
\end{equation}
where:
\begin{equation}
    \begin{aligned}
        v_b&=\sqrt{1+\frac{\Delta s^2}{8}-\sqrt{\frac{\Delta s^2}{4}+\frac{\Delta s^4}{64}}}\\
        v_c&=\sqrt{1+\frac{\Delta s^2}{8}+\sqrt{\frac{\Delta s^2}{4}+\frac{\Delta s^4}{64}}}\\
        \gamma^2&=\frac{v_b^2}{v_c^2},\quad \zeta_L=\frac{L}{l},\quad \Delta s=\Delta kl
    \end{aligned}
\end{equation}

We further investigate the noise suppression landscape by plotting the FH intensity noise reduction ratio as a joint function of wavelength detuning and input power (Fig.~\ref{fig:Fig4}a and Fig.~\ref{fig:RINTransfer}). The results reveal a distinct topology: in the phase-matched case, a single optimal operating point exists. In contrast, phase-mismatched conditions support multiple optimal noise reduction points that emerge at higher input powers. Figure~\ref{fig:RINTransfer} illustrates this complex behavior, which effectively broadens the operational bandwidth for both wavelength and pump power. 

\begin{figure}[htbp]
    \centering
    \includegraphics[width=0.5\textwidth]{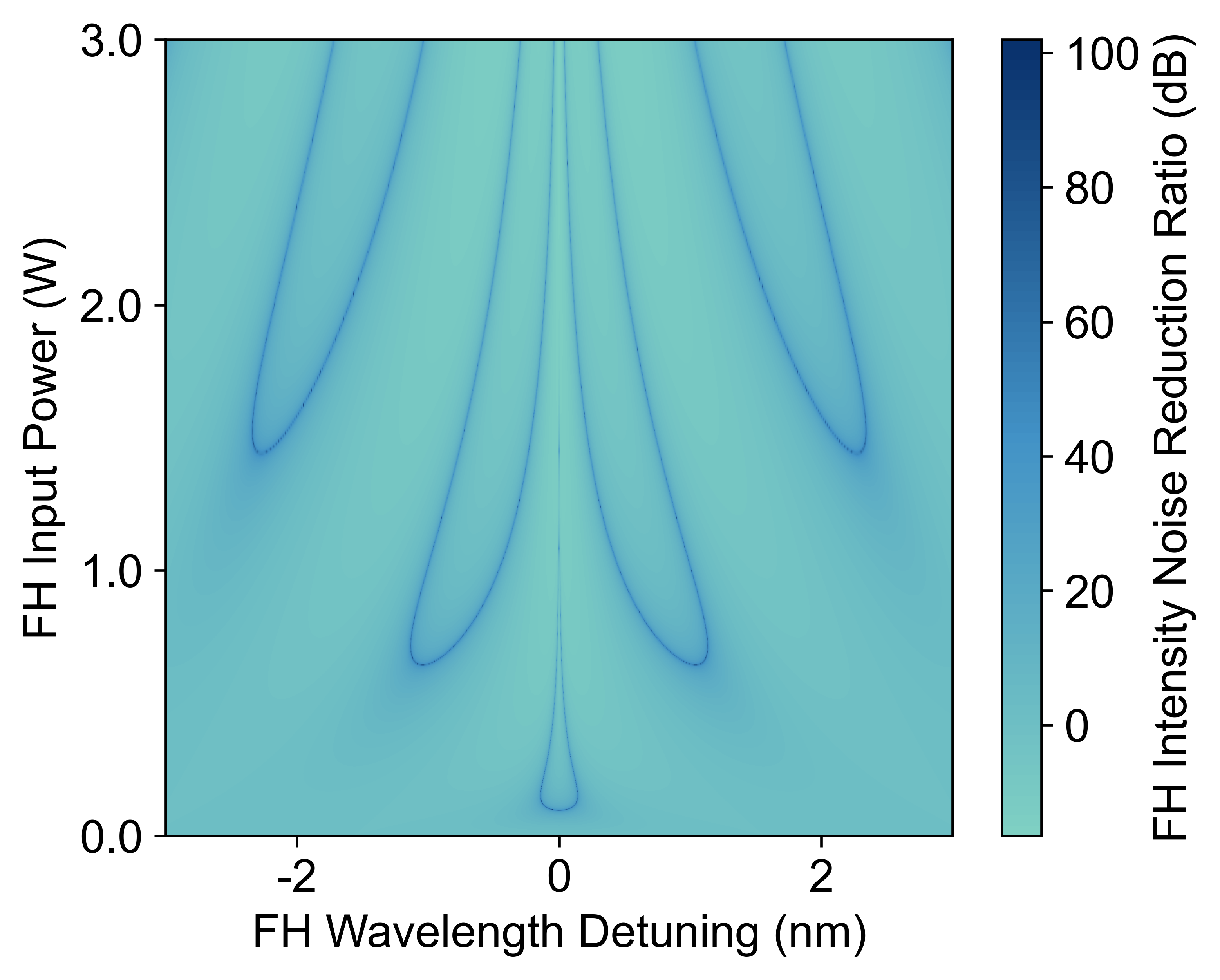}
    \caption{{\bf FH Intensity Noise Reduction Ratio versus FH Wavelength Detuning and FH Input Power.} This plot captures a broader range (0 W to 3 W) of optical power and pump wavelength detuning (-2.5 nm to 2.5 nm). Here we choose $\eta=\kappa^2=1500\%{\text{W}^{-1}}{\text{cm}^{-2}}$, $L=10$ mm, and phase-mismatch coefficient $a=-400$ rad/m/nm.}
    \label{fig:RINTransfer}
\end{figure}

Operating at these other optima requires greater power, and also leads to AM-to-PM conversion. At the QPM wavelength ($\Delta k=0$), the phase relation $\theta_p(L)=\theta_p(0)$ holds, ensuring zero AM-to-PM coupling independent of pump power. However, detuning from this condition ($\Delta k \neq 0$) breaks this symmetry, leading to power-dependent phase shifts and consequently AM-to-PM conversion. This conversion is quantified analytically in Equation~\ref{equ:analytical_sol} and plotted in Fig.~\ref{sfig:AMPMCoupling}. 

\begin{figure}[htbp]
    \centering
    \includegraphics[width=0.5\textwidth]{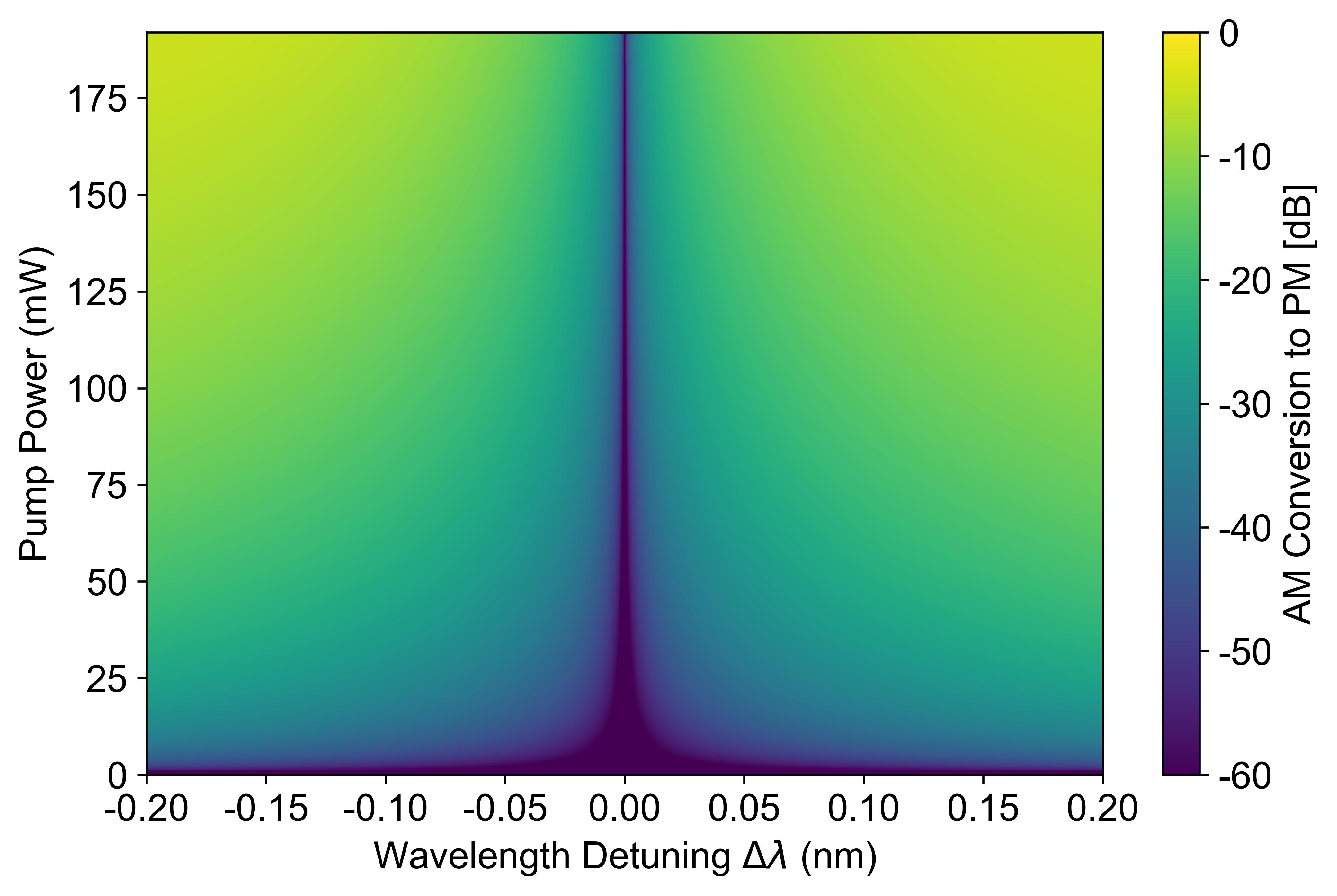}
    \caption{{\bf FH Intensity Noise Conversion to FH Phase Noise with wavelength detuning.} The AM-conversion-to-PM is defined as $|\partial \theta_p(L)/\partial\ln{P_{p,\text{in}}}|$. This simulation is done with $\eta=\kappa^2=1500\%{\text{W}^{-1}}{\text{cm}^{-2}}$, $L=10$ mm, and phase-mismatch coefficient $a=-400$ rad/m/nm.}
    \label{sfig:AMPMCoupling}
\end{figure}

\newpage
\clearpage

\section*{Poling Quality}

\begin{figure}[htbp]
\centering\includegraphics[width=1\linewidth]{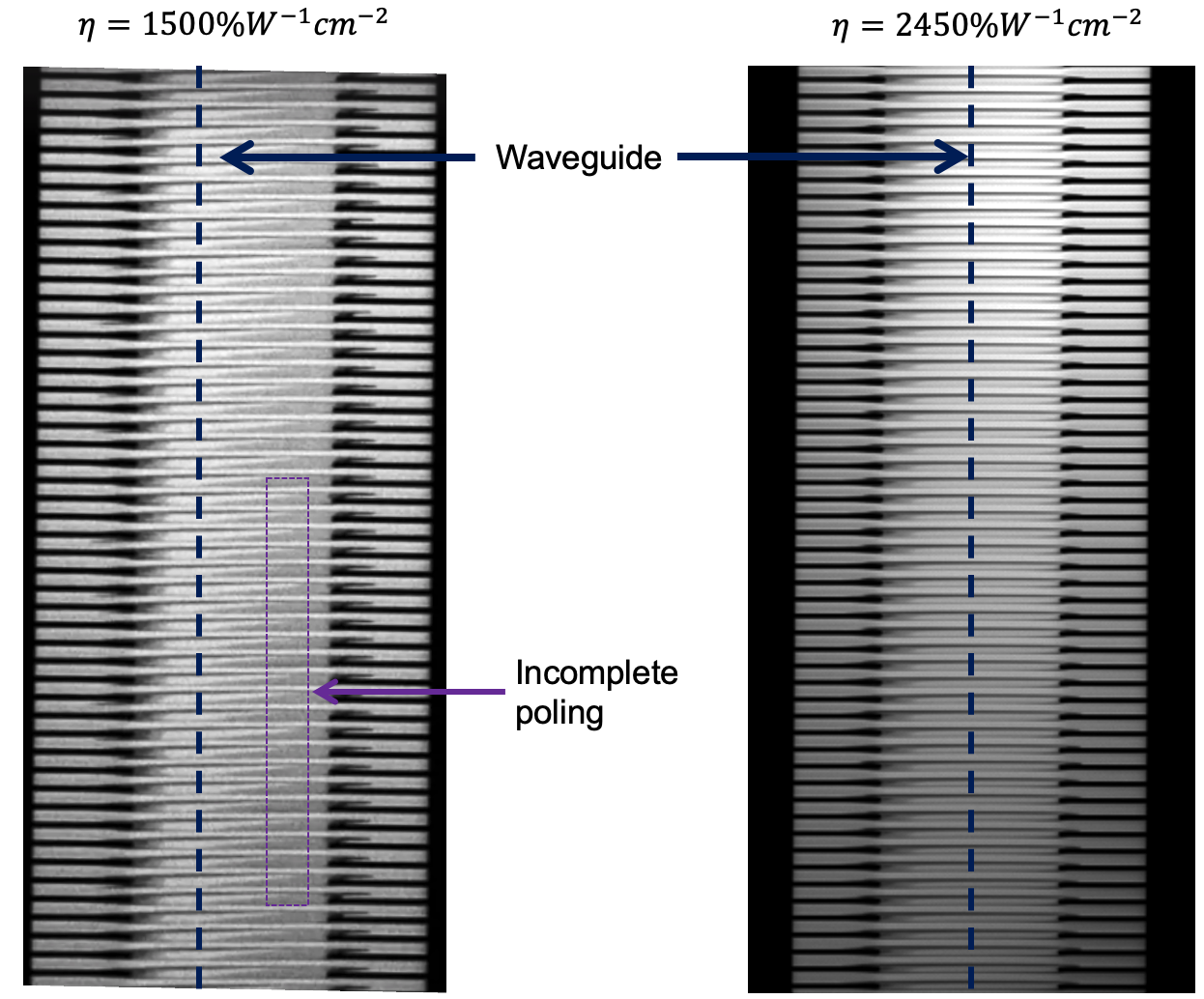}
\caption{{\bf Difference in Poling Quality of Different PINE chips.} We fabricated two PINE chips in total. We obtained most data from the first chip (left) except the length-dependent-critical-power measurement shown in Fig. \ref{fig:Fig4}c. The length dependence chip (right) has noticeably better poling quality under the second harmonic microscopy, and its $\eta$ is higher accordingly.}
\label{Sfig:poling}
\end{figure}

\newpage

\section*{Quasi-phase-matching Engineering}
In nonlinear optical applications utilizing quasi-phase-matching techniques,
it is possible to engineer the phase-mismatch along the waveguide by designing the poling period $\Lambda(z)$.
At a wavelength $\lambda$ which is close to the QPM wavelength $\lambda_{\text{QPM}}$, 
the phase-mismatch can be expressed as:
\begin{equation}
    \Delta k(\lambda;z)\approx a(\lambda-\lambda_{\text{QPM}})+\frac{2\pi}{\Lambda(z)}-\frac{2\pi}{\Lambda_{\text{QPM}}}=a(\lambda-\lambda_{\text{QPM}})+\Delta k(\lambda_{\text{QPM}};z)
\end{equation}
where:
\begin{equation}
    a=\frac{\partial \Delta k}{\partial \lambda}\Bigg|_{\lambda=\lambda_{\text{QPM}}}
\end{equation}

Thus the nonlinear wave equations become~\cite{santandrea2019general}:
    \begin{equation}
        \begin{aligned}
            \frac{\dd A}{\dd z}&=i\kappa A^*B e^{i\Phi(\lambda;z)}\\
            \frac{\dd B}{\dd z}&=i\kappa A^2 e^{-i\Phi(\lambda;z)}\\
            \Phi(\lambda;z)&=\int_0^z \Delta k(\lambda;z')\dd z'\\
        \end{aligned}
    \end{equation}

If a waveguide has $\Delta k(\lambda_{\text{QPM}};z)=0$, we call it a uniform waveguide. Otherwise, we call it a `QPM-engineered waveguide'.


\clearpage

\subsection*{Large power bandwidth}

For practical noise eater applications, a large power bandwidth is desired. 
As the pump wavelength is tuned, the input power required to achieve optimal noise reduction varies sensitively due to the wavelength-dependent phase-mismatch, as shown in Fig 4a. To study the power sensitivity of PINE, we can define the `power bandwidth' as the range of input powers over which the noise reduction ratio (NRR) remains above a certain threshold (e.g., $\text{NRR} >40 \text{dB}$).
A uniform waveguide has a different power bandwidth at different FH wavelengths. For instance, around $\lambda_{\text{FH}}-\lambda_{\text{QPM}}=\pm0.14\si{nm}$, the power bandwidth has a maximum as indicated by white dotted line in the left panel of FIG. S6. However, by introducing a linearly-varying phase-mismatch $\Delta k(\lambda_{\text{QPM}};z)$ along the waveguide as shown in FIG S5, we can achieve a large power bandwidth for noise eater applications, compared with the uniform waveguide, and their comparison at the maximum power bandwidth is shown in FIG S7. 
\begin{figure}
    \centering
    \includegraphics[width=1.0\textwidth]{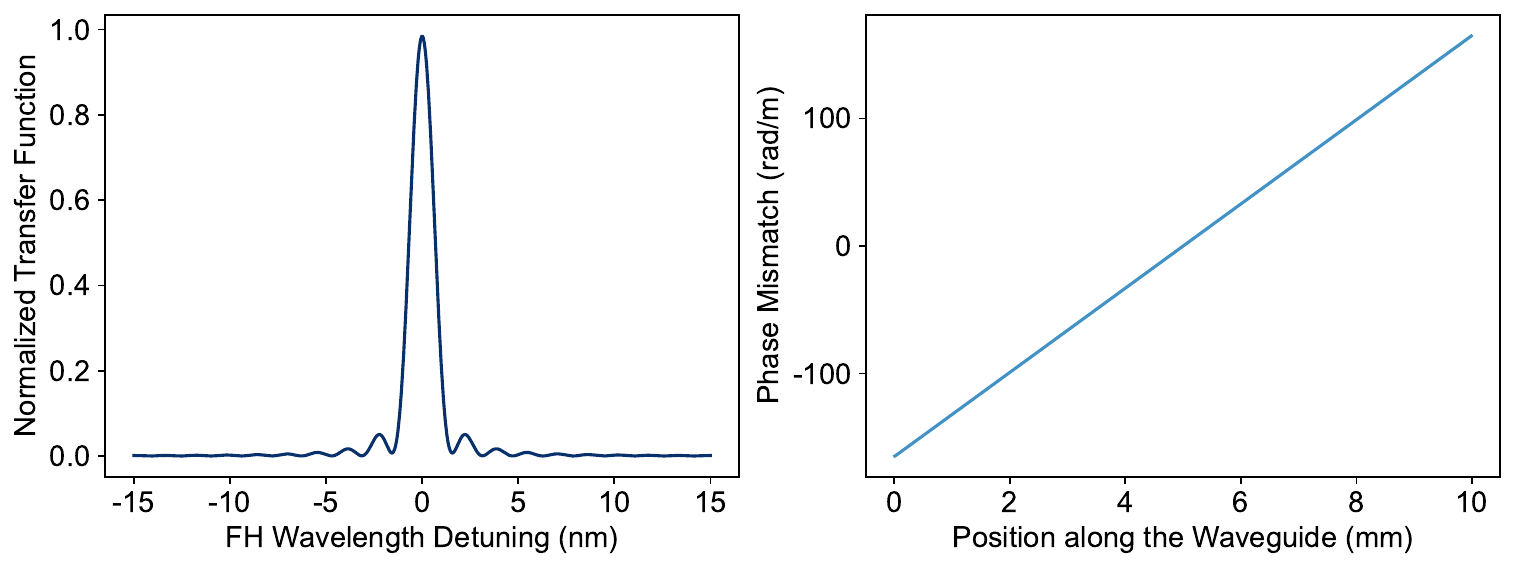}
    \caption{{\bf QPM-engineering Design (right) and Corresponding SHG Transfer Function (left) for Large Power Bandwidth Operation.}}
    \label{fig:adaptive_poling_power_output}
\end{figure}
\begin{figure}
    \centering
    \includegraphics[width=1.0\textwidth]{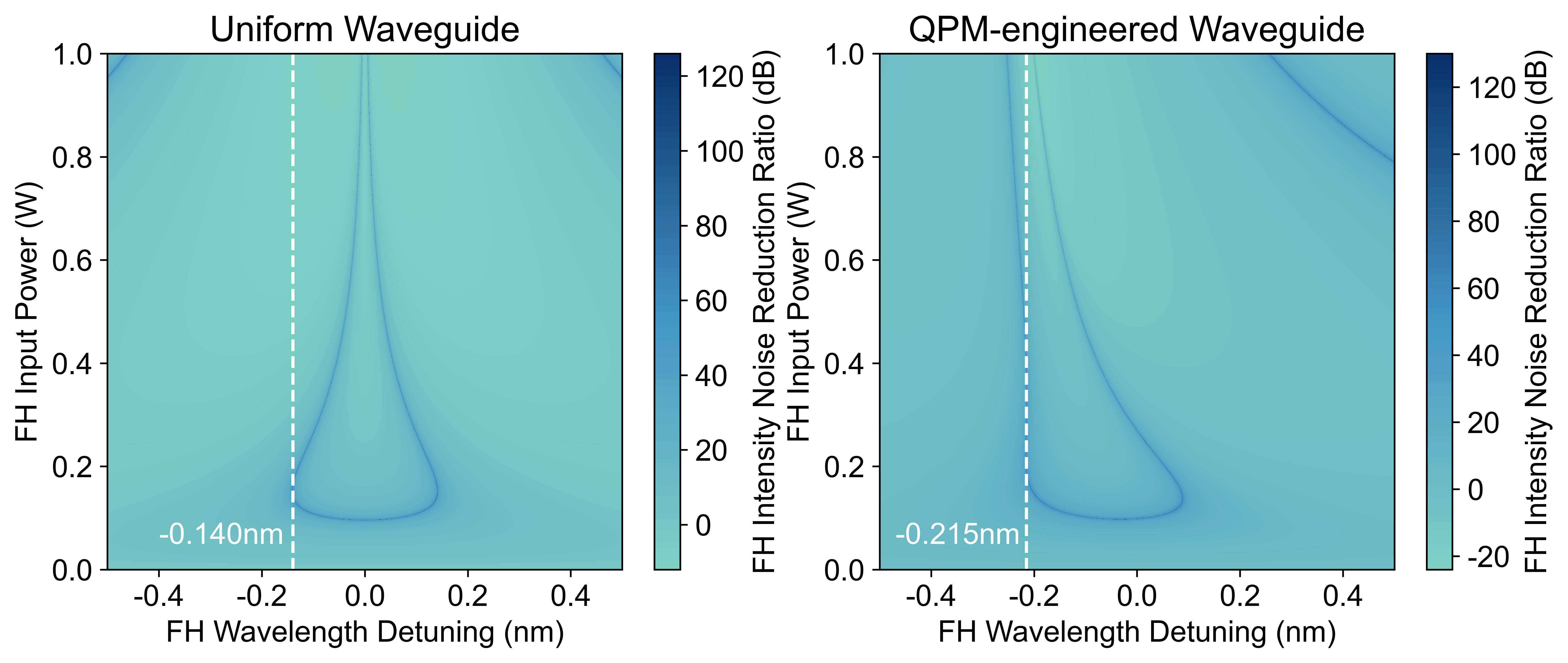}
    \caption{{\bf Intensity Noise Reduction Ratio Heatmap for Uniform Waveguide and QPM-engineered Waveguide.} The maximum power bandwidth appears at $-0.140\text{nm}$ for uniform waveguide and $-0.215\text{nm}$ for QPM-engineered waveguide.}
    \label{fig:adaptive_poling_noise_eater}
\end{figure}
\begin{figure}
    \centering
    \includegraphics[width=0.5\textwidth]{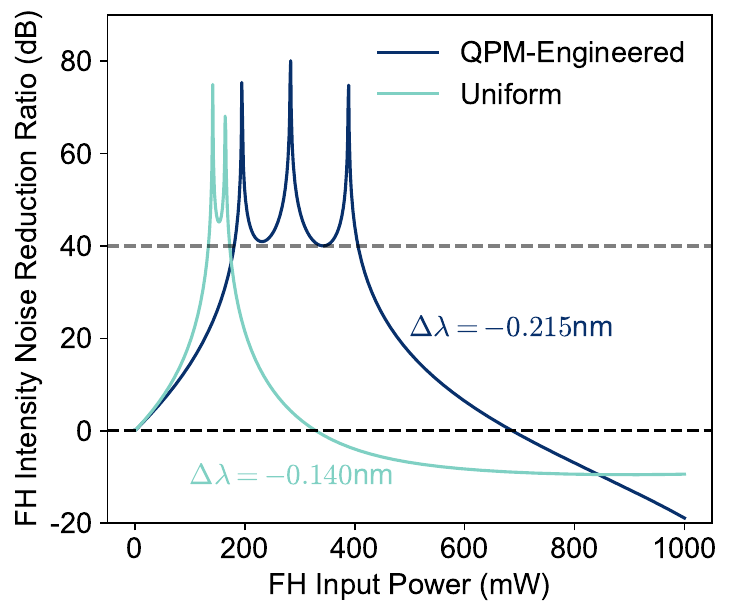}
    \caption{{\bf Comparison of FH Intensity Noise Reduction Ratio versus FH Input Power between Uniform Waveguide and QPM-engineered Waveguide.} The FH wavelength detuning is $-0.140\si{nm}$ for uniform waveguide and $-0.215\si{nm}$ for QPM-engineered waveguide.}
    \label{fig:adaptive_poling_noise_eater_vs_input_power}
\end{figure}

\subsection*{Large wavelength bandwidth}
At a specific wavelength, usually only certain input powers can achieve optimal noise reduction. We can define the `wavelength bandwidth' as the range of FH wavelengths over which the noise reduction ratio (NRR) remains above a certain threshold (e.g., $\text{NRR} >40 \text{dB}$) for a fixed input power.
We learn from the FH noise reduction heatmap in Fig. \ref{fig:Fig4}a that a uniform waveguide has a maximum wavelength bandwidth at the critical input power, corresponding to the first lobe region in the heatmap.
If the FH input power goes higher, then a second maximum wavelength bandwidth appears at two detuned wavelengths, corresponding to the second lobe regions in the heatmap.

We found that by introducing a Gaussian-like phase-mismatch $\Delta k(\lambda_{\text{QPM}};z)$ along the waveguide, we can achieve a large wavelength bandwidth for noise eater applications, compared with the uniform waveguide.
This QPM-engineered waveguide can maintain effective noise reduction over a broader range of wavelengths, making it more versatile for practical applications.
In the first lobe region of the heatmap, the wavelength bandwidth is increased by approximately 21\% compared to the uniform waveguide, although at the cost of increased required input power.
In the second lobe region of the heatmap, the wavelength bandwidth is increased by approximately 43\% compared to the uniform waveguide. 
What is better is that the required input power is actually reduced.

\begin{figure}
    \centering
    \includegraphics[width=1.0\textwidth]{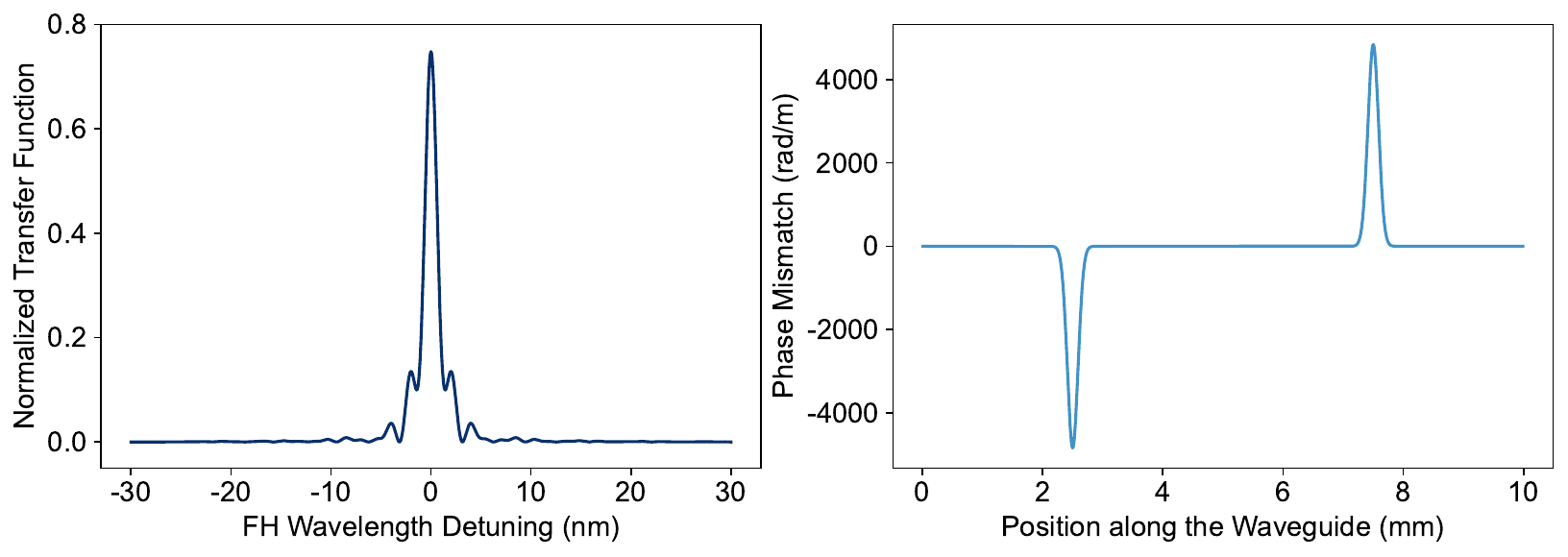}
    \caption{{\bf QPM-engineering Design (right) and Corresponding SHG Transfer Function (left) for Larger Wavelength Bandwidth Operation.}}
    \label{fig:adaptive_poling_wavelength_output}
\end{figure}

\begin{figure}
    \centering
    \includegraphics[width=1.0\textwidth]{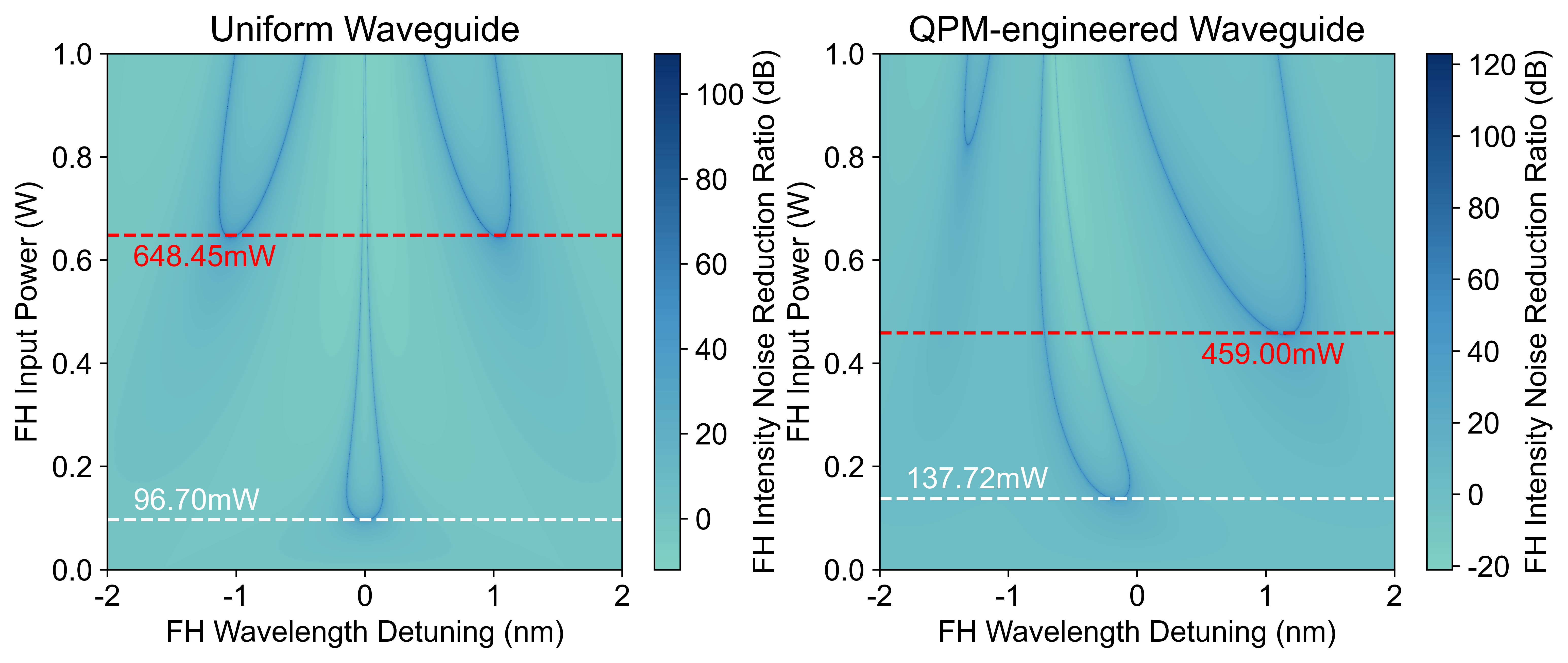}
    \caption{{\bf Intensity Noise Reduction Ratio Heatmap for Uniform Waveguide and QPM-engineered Waveguide.} On both plots we could see first order lobe whose minimum input power is indicated by white dashed lines, as well as second order lobe indicated by red dashed lines.}
    \label{fig:adaptive_poling_wavelength_noise_eater}
\end{figure}

\begin{figure}
    \centering
    \includegraphics[width=1.0\textwidth]{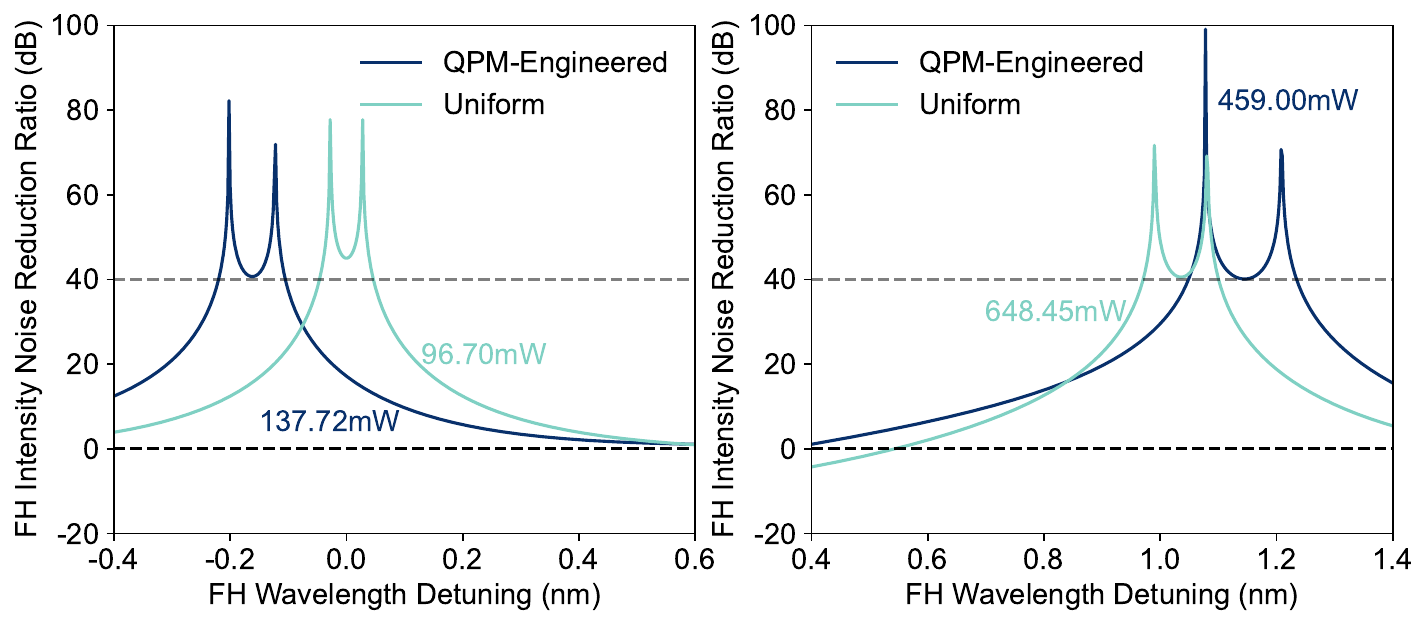}
    \caption{{\bf Comparison of FH Intensity Noise Reduction Ratio versus FH Wavelength Detuning between Uniform Waveguide and QPM-engineered Waveguide.} Left: comparing the first lobe region, $P_{\text{in}}=96.70\text{ mW}$ for uniform waveguide and $P_{\text{in}}=137.72\text{ mW}$ for QPM-engineered waveguide; Right: comparing the second lobe region, $P_{\text{in}}=648.4\text{ mW}$ for uniform waveguide and $P_{\text{in}}=459.0\text{ mW}$ for QPM-engineered waveguide.}
    \label{fig:adaptive_poling_wavelength_noise_eater_vs_wavelength}
\end{figure}



\stopcontents[post]
\end{document}